\documentclass[usenatbib]{mn2e}
\usepackage{color}
\usepackage{graphicx}
\usepackage{epsf}
\usepackage{epsfig,graphicx,psfrag,amsfonts,amssymb}
\usepackage{amsmath}
\usepackage{dcolumn}
\usepackage{bm}
\def \lleq {\lower0.9ex\hbox{ $\buildrel < \over \sim$} ~}
\def \ggeq {\lower0.9ex\hbox{ $\buildrel > \over \sim$} ~}

\newcommand{\m}{{\rm m}}

\def \beq  {\begin{equation}}
\def \eeq  {\end{equation}}
\def \ber  {\begin{eqnarray}}
\def \eer  {\end{eqnarray}}
\def\statei{\lbrace r,s \rbrace}

\newcommand{\be}{\begin{equation}}
\newcommand{\ee}{\end{equation}}
\newcommand{\ba}{\begin{eqnarray}}
\newcommand{\ea}{\end{eqnarray}}
\newcommand{\bea}{\begin{eqnarray*}}
\newcommand{\eea}{\end{eqnarray*}}

\begin{document}
\title[]
{$Om$ diagnostic applied to scalar field models and slowing down of cosmic acceleration}
\author[]
{M. Shahalam$^{1}$\thanks{mohdshahamu@gmail.com }, Sasha Sami$^{2}$\thanks{sashasami03@gmail.com},  Abhineet Agarwal$^2$\thanks{agarwal.abhi93@gmail.com}\\
$^1$Center for Theoretical Physics, Jamia Millia Islamia, New Delhi-110025, India\\
$^2$International Institute of Information Technology Gachibowli, Hyderabad 500032, Telangana, India}

\date{\today}
\maketitle
\pagerange{\pageref{firstpage}--\pageref{lastpage}} \pubyear{2010}

\begin{abstract}
We apply the $Om$ diagnostic to models for dark energy based on scalar fields.
In case of the power law potentials, we demonstrate the
 possibility of slowing down the expansion of the Universe
 around the present epoch for a specific range in the parameter space. For these models, we also examine the issues concerning the age of Universe. We use the $Om$ diagnostic to distinguish the $\Lambda$CDM model from non minimally coupled scalar field, phantom field and generic quintessence models. Our study shows that the $Om$ has zero, positive and negative curvatures for $\Lambda$CDM, phantom and quintessence models respectively. We use an integrated data base (SN+Hubble+BAO+CMB) for observational analysis and demonstrate that $Om$ is a useful diagnostic to apply to observational data.

\end{abstract}

\begin{keywords}
cosmological parameters$-$ cosmology: observations$-$ cosmology: theory$-$ dark energy.
\end{keywords}

\section{Introduction}
\label{intro} A number of observational investigations such as,
Type Ia supernovae (Riess et al. 1998; Perlmutter et al. 1999),
cosmic microwave background radiation (Spergel et al. 2003; Komatsu et al. 2011), surveys of the large scale structure (Eisenstein et al. 2005) and PLANCK 2013 (Ade et al. 2013) indicate that our Universe is accelerating at present. In the standard framework based upon Einstein gravity, cosmic acceleration can be explained by  an exotic fluid with large negative pressure filling the Universe, dubbed `{\it dark energy'} (Riess et al. 1998; Perlmutter et al. 1999; Sahni \&  Starobinsky 2000; Spergel et al. 2003; Copeland, Sami \& Tsujikawa 2006; Sami 2009; Komatsu et al. 2011; Ade et al. 2013; Sami \&  Myrzakulov 2013). The simplest candidate for dark energy (DE)$-$ the cosmological constant $\Lambda$ is plagued with difficult theoretical issues related to fine tuning. A variety of other dark energy models have been studied in the literature, namely, quintessence (Ratra \&  Peebels 1988), $\Lambda$CDM (Weinberg 1989) and the phantom field (Caldwell, Kamionkowski \& Weinberg 2003;  Setare 2007).

   Alternatively, cosmic acceleration can also be mimicked by the large scale modification of gravity  (Boisseau et al. 2000; Dvali, Gabadadze \& Porrati 2000; Sahni \& Shtanov 2003; Sahni, Shtanov \& Viznyuk 2005; Trodden 2007; Brevik 2008; Ali, Gannouji \& Sami 2010; Gannouji \& Sami 2010). Though, the late time cosmic acceleration is considered to be an established phenomenon at present, its underlying cause remains uncertain. A large number of models, within the framework of standard lore and modified theories of gravity can explain the said phenomenon. It is therefore important to find ways to discriminate between various competing models. To this effect, important geometrical diagnostics have been recently suggested in the literature such as, Statefinder (Alam et al. 2003; Sahni et al. 2003), Statefinder hierarchy (Arabsalmani \& Sahni 2011), $Om$ (Sahni, Shafieloo \& Starobinsky 2008, 2014) and $Om$3 (Shafieloo, Sahni \& Starobinsky 2012). Statefinders use the second, third and higher order derivatives of the scale factor with respect to cosmic time whereas $Om$ relies on first order derivative alone. Consequently $Om$ is a simpler diagnostic when applied to observations.

The statefinder method has been extensively used in the literature to distinguish among various models of dark energy and modified theories of gravity. For instance, non minimally coupled scalar field, galileon field, Dvali, Gabadadze and Porrati (DGP) model, bimetric (bigravity) theory of massive gravity and others have been investigated (Sami et al. 2012; Myrzakulov \& Shahalam 2013) using this diagnostic. Observational constraints have been put on the statefinder pair $\statei$ and deceleration parameter $q$  using Union2.1 compilation data (Suzuki et al. 2012) and 28 points of Hubble data (Farooq \& Ratra 2013), for the power law cosmological model (Rani et al. 2014). The statefinder analysis has been applied to non minimally coupled galileons and $f(T)$ cosmology by several authors (Jamil et al. 2012; Jamil, Momeni \& Myrzakulov 2013). Recently, statefinder hierarchy  has been applied to models based upon  Chaplygin gas, light mass galileons and holographic dark energy (Li, Yang \& Chen 2014; Myrzakulov \& Shahalam 2014; Zhang, Cui \& Zhang 2014).

Coming back to $Om$ diagnostic, to be employed in this paper, we should note its excellent features. First, it can discriminate dynamical dark
 energy models from $\Lambda$CDM, in a robust way, even if the
 value of the matter density is not precisely known. Secondly,
 it can provide a null test of $\Lambda$CDM hypothesis, i.e., $Om(z)$ - $\Omega_{0m}$=0, if dark energy is a cosmological constant. $Om$ has zero, negative and positive curvatures for $\Lambda$CDM, quintessence and phantom models respectively.

Our paper is organised as follows: In section \ref{om diagnostic}, we briefly revisit the $Om$ diagnostic, to be used in the subsequent sections.
   In section \ref{EOM}, we  display evolution equations of  phantom (non-phantom) scalar field in the autonomous form. In sub-section \ref{quint},
we consider power law potentials and examine the slowing down of cosmic acceleration for a possible range of the parametric space, whereas in sub-section \ref{cosmicage} we consider age of the Universe in these models. In sub-section \ref{tracker}, we examine equation of state (EOS) and $Om$ behaviour for the tracking potential $V(\phi)=V_0\left[\cosh ({\tilde \alpha} \phi/M_p)-1\right]^p$. Sub-section \ref{phantom} and section \ref{NMC} are devoted to application  of $Om$ to phantom field with linear potential and non minimally coupled scalar field respectively. Our results are summarized in the last section. In appendix A, we carry out joint data analysis and put observational constraints on the model parameters using Union2.1 compilation data (Suzuki et al. 2012), 28 points
of Hubble data (Farooq \& Ratra 2013), BAO data (Eisenstein et al. 2005; Percival et al. 2010; Beutler et al. 2011; Blake et al. 2011; Jarosik et al. 2011; Giostri et al. 2012) and CMB data (Komatsu et al. 2011).
\section{$O{\m}$ diagnostic}
\label{om diagnostic}
\begin{figure}
\begin{center}
\begin{tabular}{c }
{\includegraphics[width=2.5in,height=2.5in,angle=0]{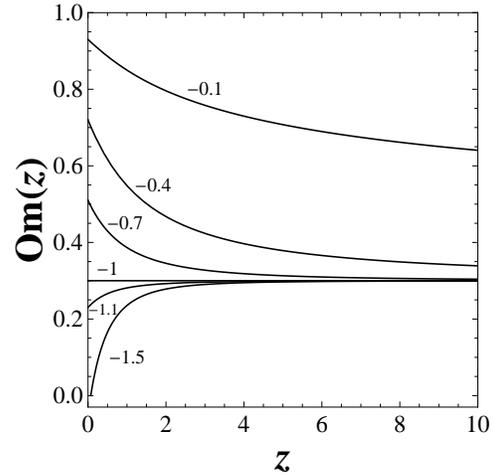}}
\end{tabular}
\caption{ This figure shows the evolution of $Om(z)$ versus the redshift $z$ for DE models with $\Omega_{0m} = 0.3$ and $w$ = -1.5, -1.1, -1, -0.7, -0.4, -0.1 (lower to top). The horizontal line represents $\Lambda$CDM with $w$ = -1, and has zero curvature. The DE models with $w > -1$ (quintessence) have negative curvature whereas DE models with $w < -1$ (phantom) have positive curvature.}
\label{wslowd}
\end{center}
\end{figure}

$Om$ is a geometrical diagnostic which
combines Hubble parameter and redshift. It can differentiate a
dynamical dark energy model from $\Lambda$CDM, with and without
reference to matter density. Constant behaviour of $Om(z)$ with
respect to $z$  signifies that DE is a cosmological constant
($\Lambda$CDM). The positive slope of $Om(z)$ implies that dark
energy is phantom ($w < -1$) whereas the negative slope means that DE behaves like quintessence ($w > -1$). Following Sahni et al. (2008); Zunckel \& Clarkson (2008), $Om(z)$ for spatially flat Universe is defined as
\be
Om(z) \equiv \frac{H^2(z)/{H_0^2}-1}{(1+z)^3-1},
\label{eq:om} \ee
where $H_0$ is the present value of the Hubble
parameter. Since, for constant equation of state parameter,
 \be
H^2(z) = H_0^2(\Omega_{0m}(1+z)^3 + (1-\Omega_{0m})(1+z)^{3(1+w)}),
\label{eq:hubble}
 \ee
 we have the following expression for $Om(z)$,
\be Om(z) = \Omega_{0m} + (1-\Omega_{0m})\frac{(1+z)^{3(1+w)}-
1}{(1+z)^3-1}~,
\ee
which shows that, $Om(z) = \Omega_{0m}$ for
$\Lambda$CDM, whereas $Om(z) > \Omega_{0m}$ for quintessence $(w
> -1)$ and $Om(z) < \Omega_{0m}$ for phantom $(w < -1)$. This kind
of behaviour can be clearly seen in figure \ref{wslowd}. It is
remarkable  that $Om$ provides a null test of the $\Lambda$CDM
hypothesis. Another conclusion, that follows from this figure is that
the growth  of $Om(z)$ at late time favours the decaying dark
energy models (Shafieloo, Sahni \& Starobinsky 2009).
\begin{figure*} \centering
\begin{center}
$\begin{array}{c@{\hspace{0.2in}}c c}
\multicolumn{1}{l}{\mbox{}} &
        \multicolumn{1}{l}{\mbox{}} \\ [0.0cm]
\epsfxsize=2in
\epsffile{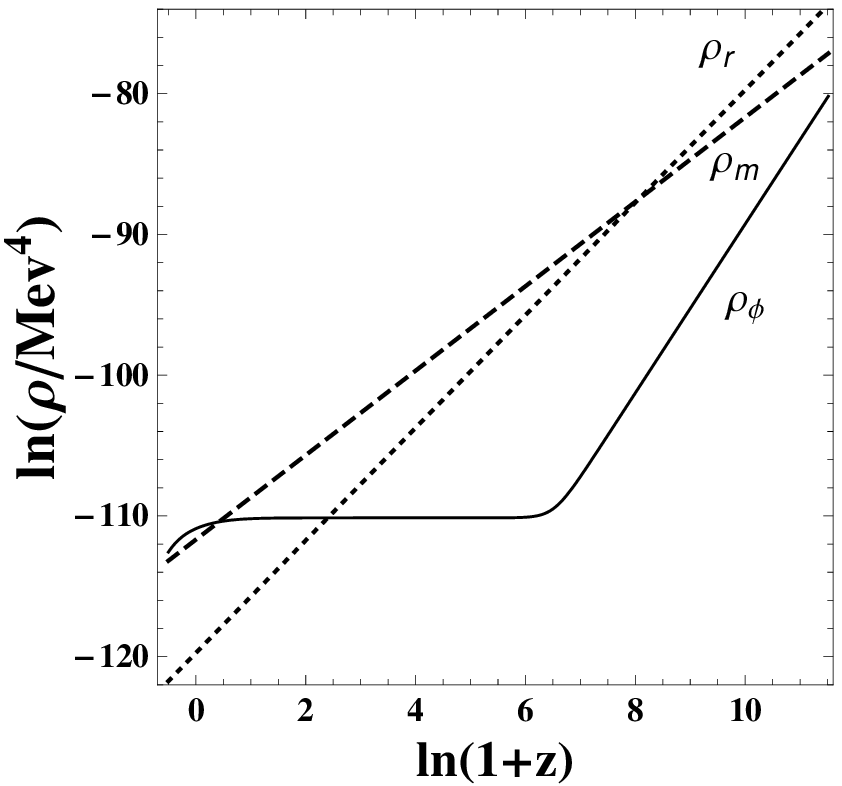} &
        \epsfxsize=1.91in
        \epsffile{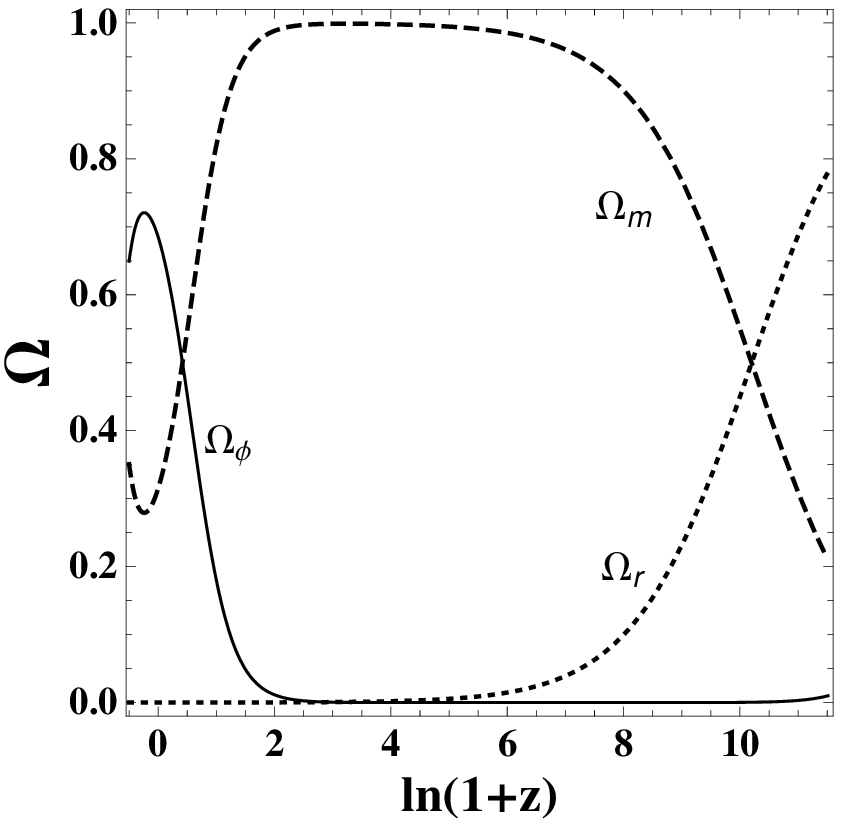} &
  \epsfxsize=2.08in
        \epsffile{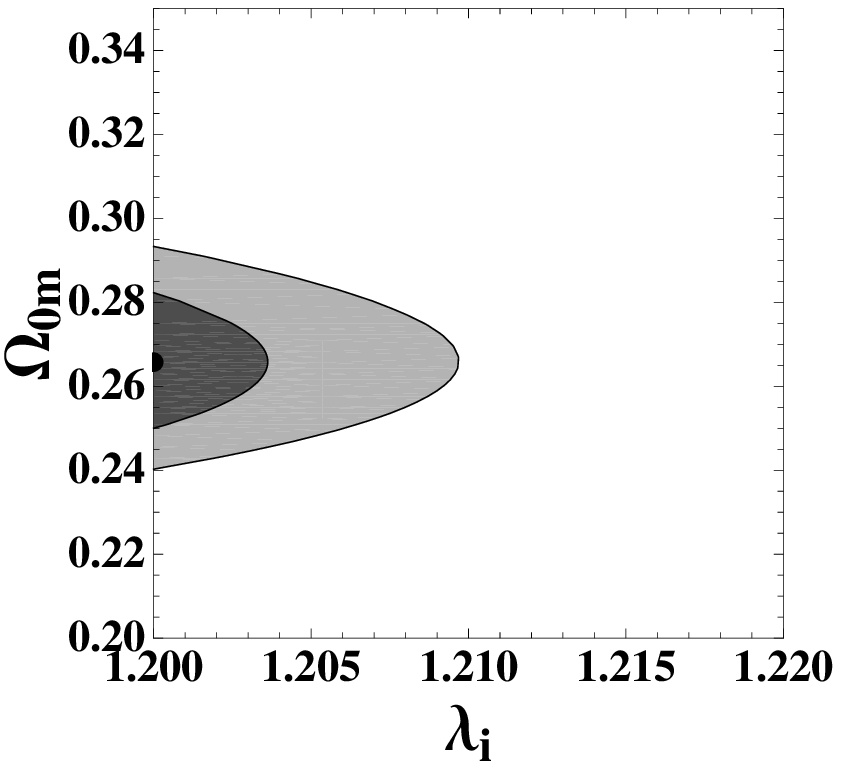}
\\
\multicolumn{1}{l}{\mbox{}} &
        \multicolumn{1}{l}{\mbox{}} \\ [0.0cm]
\epsfxsize=2in
\epsffile{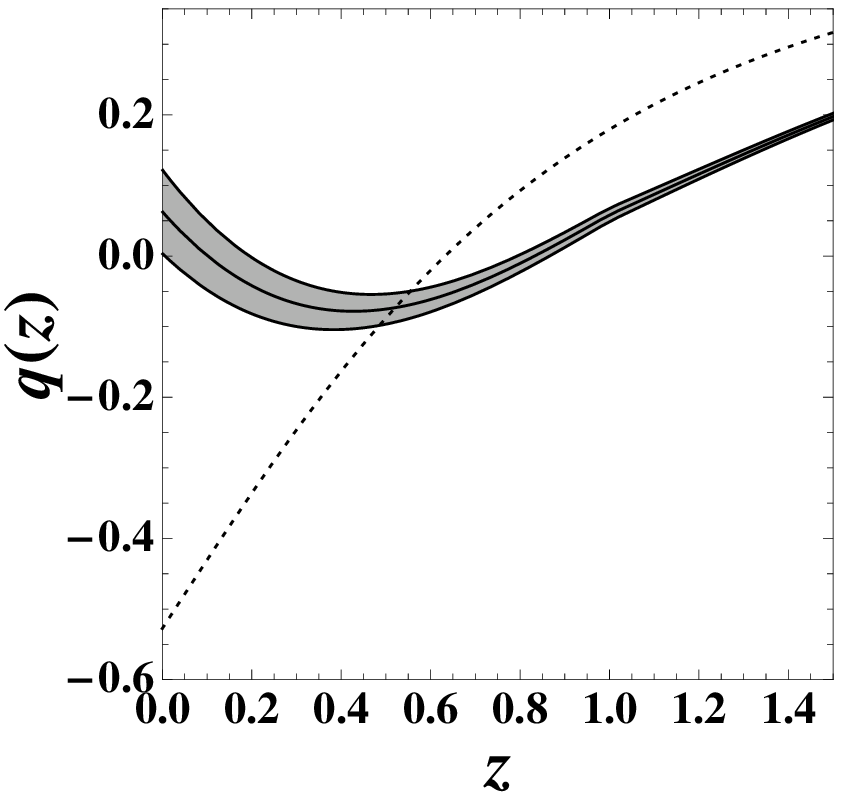} &
        \epsfxsize=2in
        \epsffile{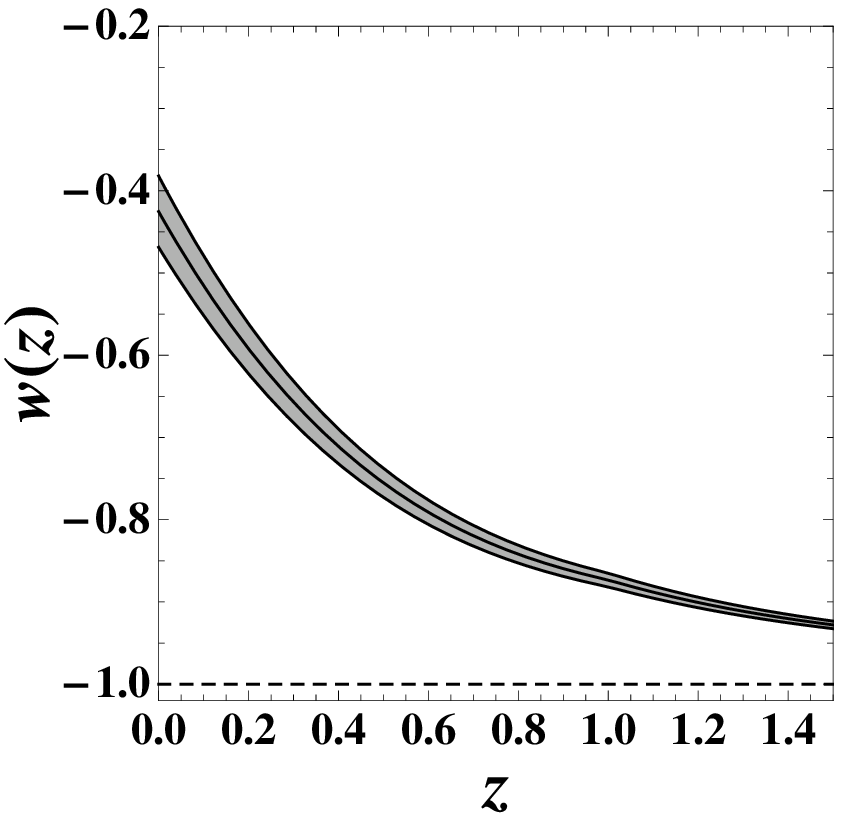} &
  \epsfxsize=1.97in
        \epsffile{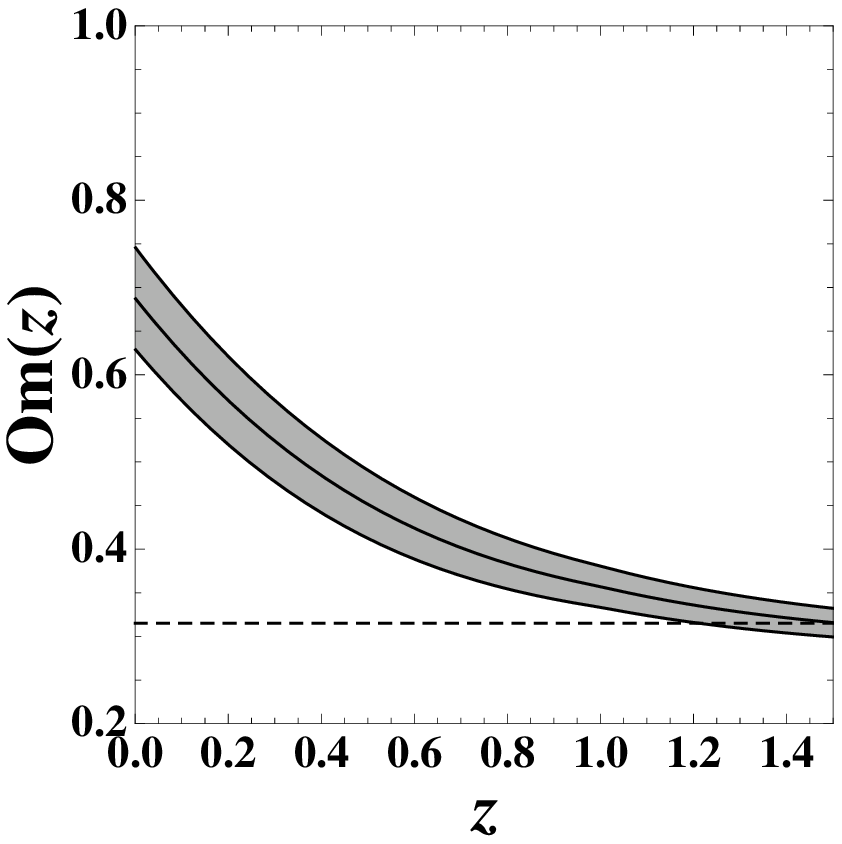}

\end{array}$
\end{center}
\caption{ This figure shows different plots for the potential (\ref{eq:phi2pot}) with $m=1$. The upper
left plot shows the evolution of energy density versus the redshift $z$. The dotted,
dashed and black lines correspond to 
 the energy density of radiation, matter and scalar field respectively. Initially, the
 energy density of scalar field ($\rho_\phi$) is extremely sub-dominant and remains so, for most of the period of evolution. At late times, the field energy density catches up with the background
 and overtakes it. Around  the present epoch, the field energy density starts decaying and correspondingly its equation of state starts approaching towards matter. The upper middle plot shows the
 evolution of density parameter ($\Omega$) versus the redshift $z$. 
The upper right plot shows the 1$\sigma$ (dark shaded) and 2$\sigma$ (light shaded)
likelihood contours in the $\lambda_i - \Omega_{0m}$ plane, where $\lambda_i$ is the initial value of $\lambda$. The black dot designates the best fit value. The lower left plot shows the evolution of $q$ versus the redshift $z$. The value of $q$ increases at redshift $z\lesssim 0.4$ and becomes positive at late times.  The lower middle and lower right  plots show the evolution of $w(z)$ and Om(z) versus the redshift $z$. In all lower plots, dashed line represents $\Lambda$CDM with $\Omega_{0m}=0.315$; solid (middle) line inside shaded regions show best fitted behaviour and shaded regions show 1$\sigma$ confidence level. We have used joint data (SN+Hubble+BAO+CMB), see appendix A for details.}
\label{figphi2}
\end{figure*}
\begin{figure*} \centering
\begin{center}
$\begin{array}{c@{\hspace{0.2in}}c c}
\multicolumn{1}{l}{\mbox{}} &
        \multicolumn{1}{l}{\mbox{}} \\ [0.0cm]
\epsfxsize=2in
\epsffile{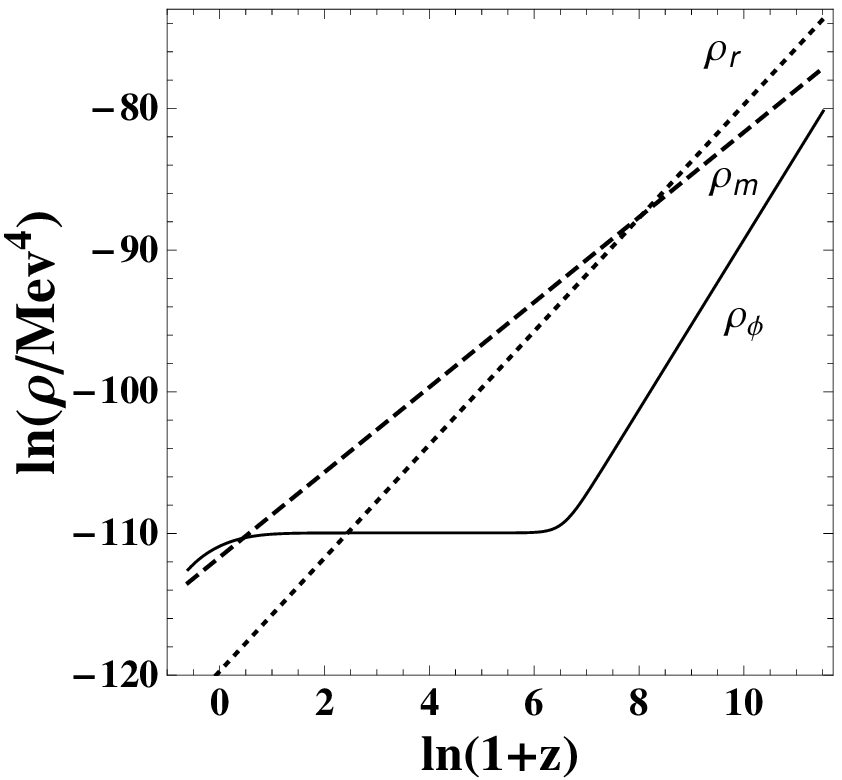} &
        \epsfxsize=1.91in
        \epsffile{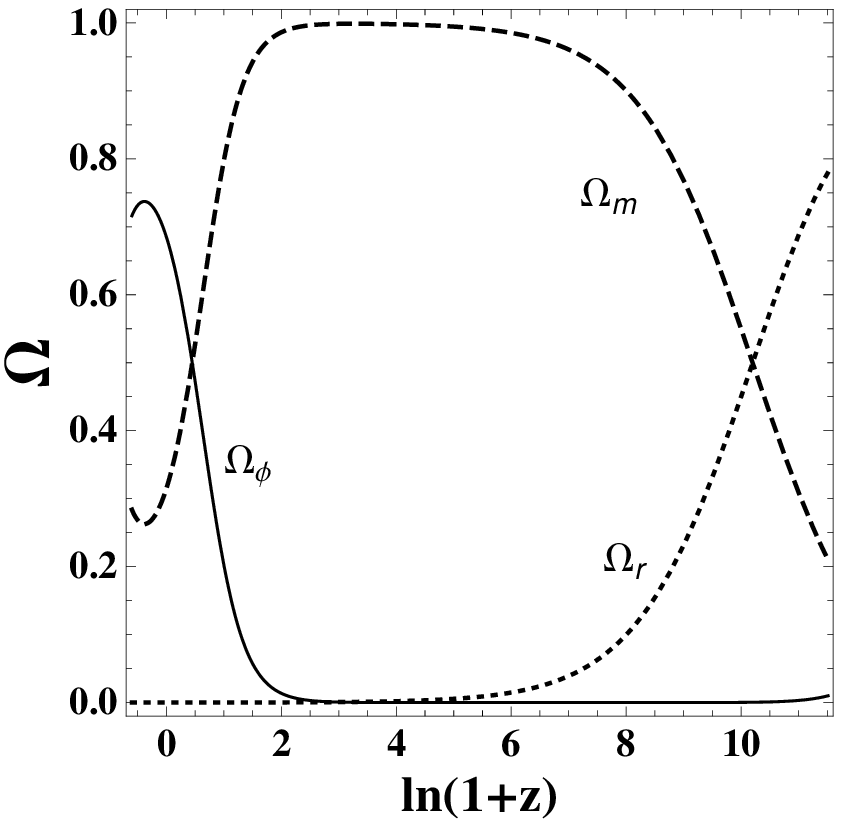} &
  \epsfxsize=2.08in
        \epsffile{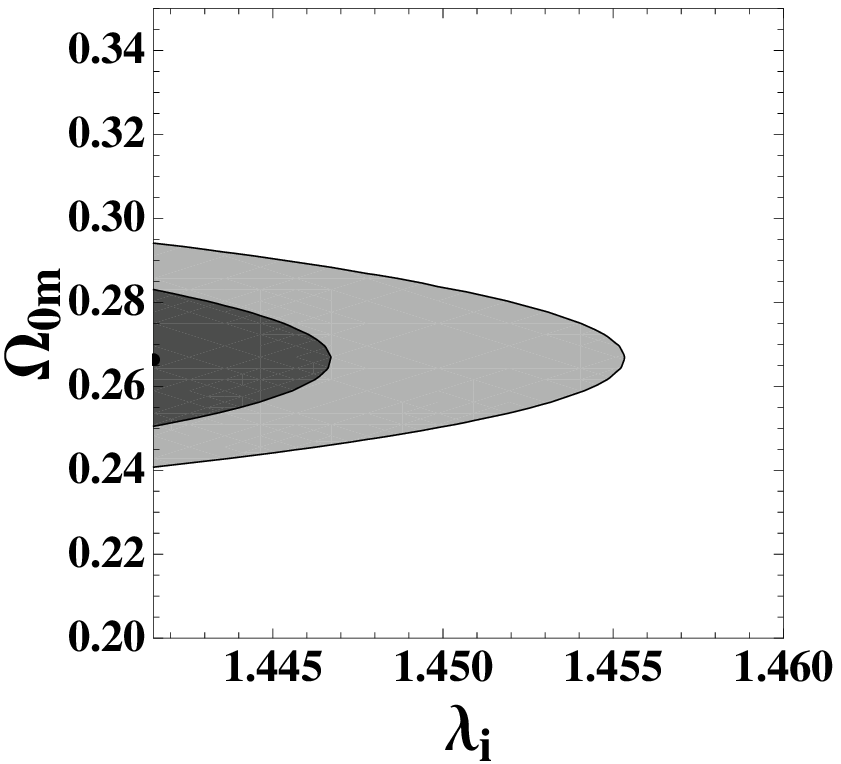}
\\
\multicolumn{1}{l}{\mbox{}} &
        \multicolumn{1}{l}{\mbox{}} \\ [0.0cm]
\epsfxsize=2in
\epsffile{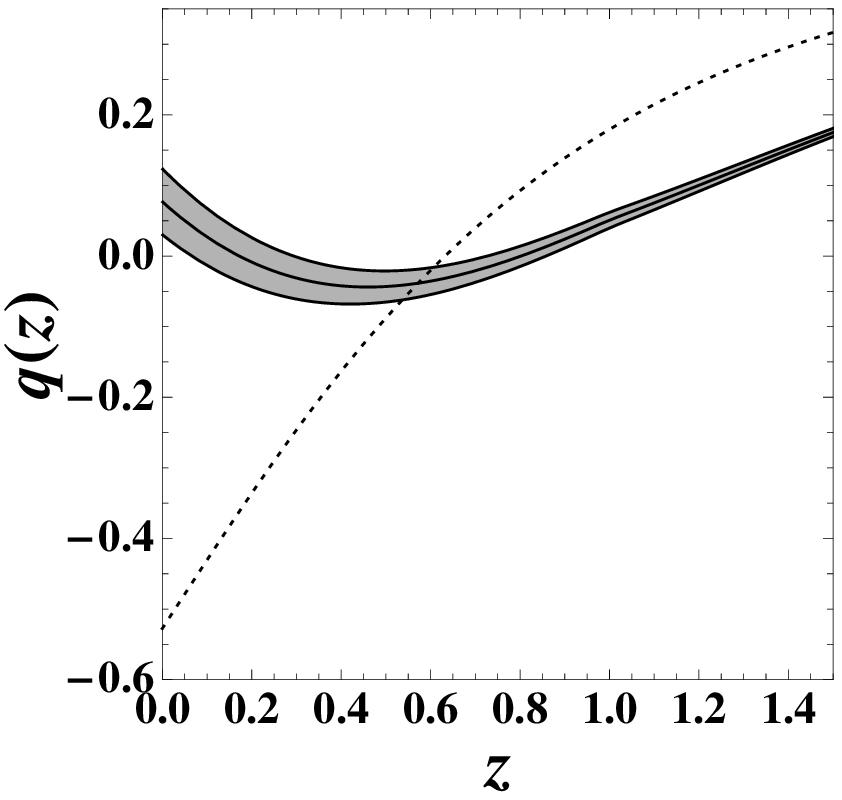} &
        \epsfxsize=2in
        \epsffile{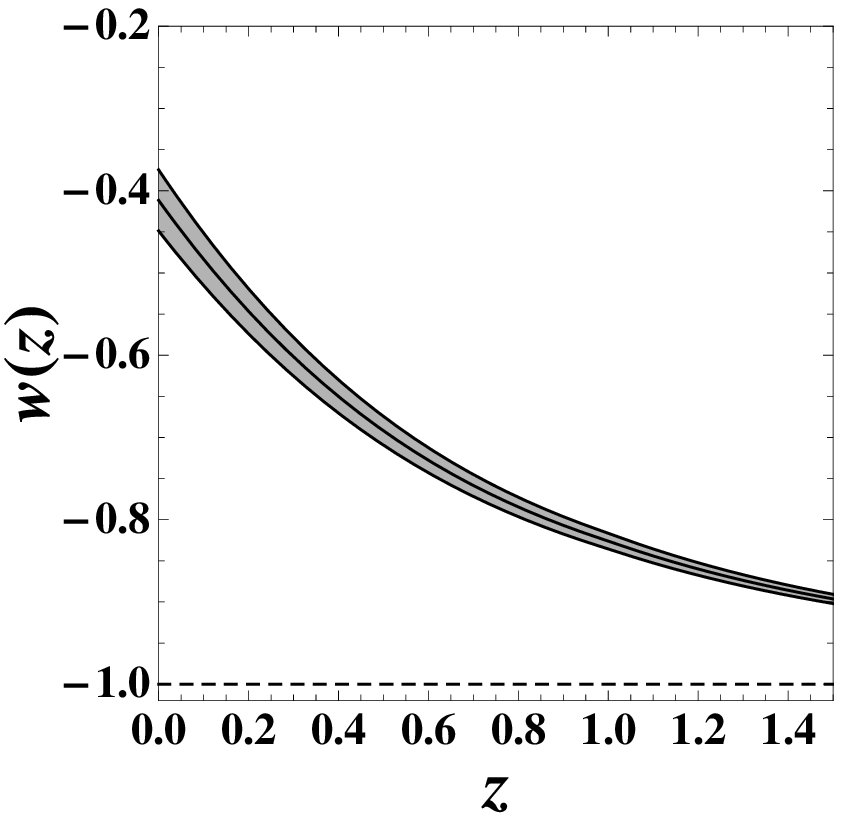} &
  \epsfxsize=1.97in
        \epsffile{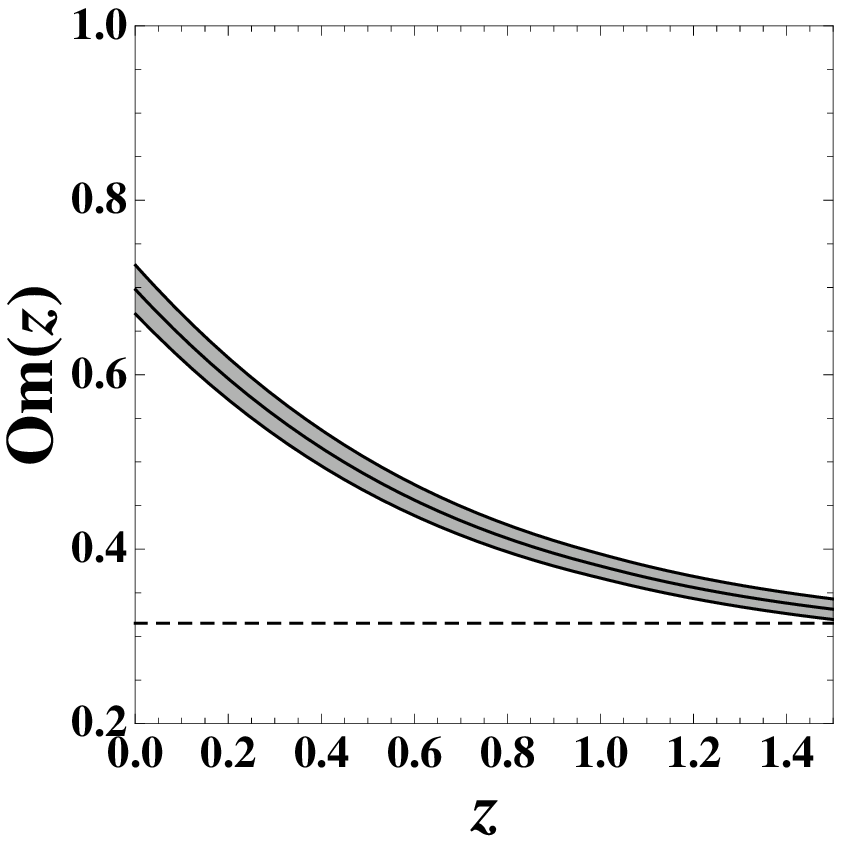}

\end{array}$
\end{center}
\caption{ This figure is similar to figure \ref{figphi2} for the potential  (\ref{eq:phi2pot}) but with $m=2$. The best fit values of the parameters are, $\lambda_i = 1.4415$ and $\Omega_{0m} = 0.2663$, where $\lambda_i$ is the initial value of $\lambda$.}
\label{figphi4}
\end{figure*}
\section {Scalar field dynamics}
\label{EOM}
In this section, we will examine cosmological dynamics
of phantom (non-phantom) field with generic potentials, using $Om$
diagnostic. We will be interested in distinguishing features
of the dynamics which allow us to differentiate these models from
$\Lambda$CDM. Secondly, we will investigate the possibility of
slowing down of cosmic acceleration in these models. The models
showing transient acceleration have been studied previously (Felder
et al. 2002; Kallosh et al. 2002; Alam, Sahni \& Starobinsky 2003; Frampton \& Takahashi 2003; Kallosh \& Linde 2003; Sahni \& Shtanov 2003).

In a spatially flat Freidmann$-$ Lemaitre$-$ Robertson$-$ Walker (FLRW) background, the equations of motion  for a scalar field take the form
\begin{align}
3M_{\rm{pl}}^2H^2 &=\rho_m+\rho_r+\rho_\phi\,,\\
M_{\rm{pl}}^2(2\dot H + 3H^2)&=-\frac{\rho_r}{3}-p_\phi\,,\\
\epsilon \ddot{\phi}+ 3H \epsilon \dot{\phi}+\frac{dV(\phi)}{d\phi} &=0,
\label{eq:phidd}
\end{align}
where,
\begin{align}
\rho_\phi &=\frac{1}{2}\epsilon \dot \phi^2 + V(\phi)\,,\nonumber\\
p_\phi &=\frac{1}{2}\epsilon \dot \phi^2 - V(\phi);~~~~ w_{\phi}=p_{\phi}/\rho_{\phi},
\label{eq:rhop}
\end{align}
and $\epsilon$ = +1 and -1 corresponds to ordinary scalar and
phantom field respectively.

Let us introduce the following dimensionless quantities,
\begin{align}
x&=\frac{\dot{\phi}}{\sqrt{6}H M_{\rm{pl}}}\,,\quad y=\frac{\sqrt{V}}{\sqrt{3} H M_{\rm{pl}}}\,, \quad \lambda=-M_{\rm{pl}}\frac{V'}{V},
\label{xy}
\end{align}
which are used to form an autonomous system,
\begin{align}
x' &=x\Bigl(\frac{\ddot{\phi}}{H\dot{\phi}}-\frac{\dot H}{H^2}\Bigr),\\
y' &=-y \Bigl(\sqrt{\frac{3}{2}}\lambda x+\frac{\dot H}{H^2}\Bigr),\\
\Omega'_r &=-2\Omega_r\Bigl(2+\frac{\dot H}{H^2}\Bigr),\\
\lambda' &=\sqrt{6}x\lambda^2(1-\Gamma),
\end{align}
where,
\begin{align}
\frac{\dot H}{H^2}&=\frac{-3-3\epsilon x^2+3 y^2-\Omega_r }{2},\\
\frac{\ddot{\phi}}{H\dot{\phi}}&=\frac{-6x+ \sqrt{6} \epsilon y^2
\lambda}{2x},\\
\Gamma & \equiv \frac{VV_{,\phi\phi}}{V_{,\phi}^2},
\end{align}
and prime ($'$) denotes the derivative with respect to $\ln a$.
\subsection{Quintessence field and slowing down of cosmic acceleration}
\label{quint}
In this sub-section, we shall consider scalar field models which can facilitate the slowing down of
cosmic acceleration, at late times. Figure \ref{wslowd} shows that $Om(z)$
 increases with increasing constant equation of state, at late times.
 This effect could correspond to dark energy decaying into dark matter or something else.
 To check this possibility, let us consider the following power law potentials,
\be 
V(\phi)=V_0 \phi ^ {2m},~m=1,2
\label{eq:phi2pot}
\ee
In this case, we have slow roll regime followed by a fast roll and the
oscillatory phase thereafter as $\phi$ approaches the origin. The average equation of state during
oscillations is given by $<w>=(m-1)/(m+1)$ and therefore, $<w>=0,1/3$ for
$m=1,2$ respectively. In this case, if we set parameters such that evolution of
Universe corresponds to slow roll regime in the recent past, then
acceleration would register its slow down around the present epoch.
Keeping this in mind, we evolve the equations of motion numerically.
Our results are displayed in figures \ref{figphi2} and
\ref{figphi4}. In figure \ref{figphi2}, the upper right plot shows
the 1$\sigma$ (dark shaded) and 2$\sigma$ (light shaded) likelihood
contours in the $\lambda_i - \Omega_{0m}$ plane, where
$\lambda_i$ is the initial value of $\lambda$. The best fit values
of the parameters are, $\lambda_i = 1.2$ and $\Omega_{0m} = 0.2657$.
The lower middle and right plots show the evolution of $w(z)$ and
$Om(z)$ versus the redshift $z$ respectively. We find that the EOS
of dark energy grows at late times  which is clearly shown by $Om$
and $q$ plots. This corresponds to an increase of $Om$ and $q$ at
redshift $z\lesssim 0.4$. We have used, SN+Hubble+BAO+CMB joint data for
analysis (see appendix A). In the parametric space, shown in $\lambda_i - \Omega_{0m}$
plane, we find that acceleration begins to slow down around $z\simeq
0.4$ followed by deceleration at present epoch (see  the lower
plots of figures \ref{figphi2} and \ref{figphi4}). We also note from
$Om$ plot that $Om(z)$ has negative curvature. The negative
curvature of quintessence allows us to distinguish  it from
$\Lambda$CDM as well as from phantom dark energy irrespective of a
given current value of the matter density.
\subsection{Slowing down of late time cosmic acceleration and age
consideration}
\label{cosmicage}
\begin{figure}
\begin{tabular}{c}
\includegraphics[width=2.5in,height=2.5in,angle=0]{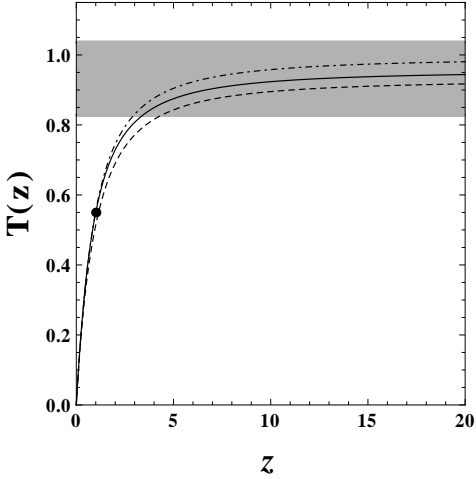}
\end{tabular}
\caption{ The dimensionless $T(z)=H_{0}t(z)$ ($t(z)$ is the cosmic age at redshift $z$) is plotted for three models, namely, $\Lambda$CDM (solid line), quintessence with $\phi^2$ potential (dashed line) and the cosmological model with given equation of state, $w(z)=-\frac{1+tanh[(z-z_t)\Delta]}{2}$ (dot dashed line) which has same number of free parameters as the CPL ansatz but does not permit the crossing to phantom divide at $w=-1$. The form of $w(z)$ has been used (Shafieloo et al. 2009) to show slowing down of cosmic acceleration at $z \lesssim 0.3$. The best fit value of $\Omega_{0m}=0.3086$ for $\Lambda$CDM, is taken from Planck data result 2013, and the best fit values of the model parameters of $\phi^2$ potential are found to be $\lambda_i=1.2$ and $\Omega_{0m}=0.2657$, (where $\lambda_i$ is the initial value of $\lambda$) whereas the best fit values of parameters for the ansatz are, $z_t=0.008$, $\Delta=12.8$ and $\Omega_{0m}=0.255$ (Shafieloo et al. 2009). The horizontal band represents the range of the age of Universe based upon observations on globular clusters (Krauss \& Chaboyer 2003; Frieman, Turner \& Huterer 2008). The black dot designates the numerical value of $T(z)$ at $z=1$ ($T(z=1)=0.5498$).}
\label{figage}
\end{figure}
The role of gravity in the Universe filled with standard matter is to decelerate the expansion such that larger is the matter density, faster will be the expansion rate thereby smaller would be 
the age of Universe. In the standard lore, the only way to circumvent the age problem is provided by invoking a repulsive effect which could be described by a positive cosmological constant or by a slowly rolling scalar field. The repulsive effect becomes important at late times giving rise to
large contribution to the age of Universe. It is therefore  not
surprising that more than half of the contribution to the  age
of Universe comes from the redshift  in the interval $z\in (0,1)$,
see figure \ref{figage}. The slowing down of acceleration also takes
place in the said interval for chosen set of model parameters. This
should clearly decrease the age of Universe. Then the requirement of
consistency with data can only be achieved by adjusting the matter
density parameter $\Omega_{0m}$ which in a sense quantifies the attractive
effect of gravity. Indeed, in case of $\phi^2$ potential, for which
slowing down commences around $z\simeq 0.4$, the best fit value of the
matter density parameter is, $\Omega_{0m}=0.2657$ which is smaller than its
$\Lambda$CDM value quoted by the Planck (see Table \ref{tabage}). Let
us also  note that the best fit value of matter density parameter
for the model ($w(z)=-\frac{1+tanh[(z-z_t)\Delta]}{2}$) given by
Shafieloo et. al (2009), is smaller than its counter part obtained
in $\phi^2$ model giving rise to some what larger value of the age 
(see figure \ref{figage}).

The cosmic age $t(z)$ is defined as the cosmic time, Universe spends from a given  $z$ to the present epoch as
\begin{equation}
t(z)=\int_{0}^{z} \frac{dz'}{(1+z')H(z')},
\label{tage}
\end{equation}
where $H(z)$ is the Hubble parameter expressed in terms of redshift
$z$. For spatially flat Universe, Hubble parameter can be written as
\begin{table}
\caption{Age summary.}
\label{tabage}
\begin{center}
\begin{tabular}{cccc}
\hline\hline
Data/Model& $\Omega_{0m}$& $H_0$& $t_0$\\
          &              & $(Km/s/Mpc)$& $(by)$\\
\hline
Planck &  0.3086~ & 67.77~& 13.7965
\\
\hline
$\Lambda$CDM &  0.3086~ & 67.77~& 13.8567\\
\hline
$\phi^2$ potential &  0.2657~ & 67.77~& 13.4825\\
\hline
$w(z)=-\frac{1+tanh[(z-z_t)\Delta]}{2}$~ &  0.255~ & 67.77~& 14.4039\\
\hline\hline
\end{tabular}
\end{center}
\end{table}
\begin{eqnarray}
H(z)&=&H_0 \Big{[} \Omega_{0m} (1+z)^3 +  (1-\Omega_{0m})\nonumber\\
&& \exp \left( 3 \int_{0}^{z} \left[ 1 + w(z') \right]
\frac{dz'}{1+z'} \right) \Big{]}^{1/2},
\label{Hage}
\end{eqnarray}
where $w=-1$ for $\Lambda$CDM, $w(z)=\frac{\frac{1}{2}\dot{\phi}^2-V(\phi)}{\frac{1}{2}\dot{\phi}^2+V(\phi)}$ for
 quintessence with $\phi^2$ potential and $w(z)=-\frac{1+tanh[(z-z_t)\Delta]}{2}$ for the cosmological model
 suggested by Shafieloo et al. (2009).

It is helpful to use a dimensionless cosmic age $T (z) = H_0 t(z)$,
where $t(z)$ is given by equation (\ref{tage}) such  that $T (z) =
H_0 t(z) \rightarrow 0$ as $z \rightarrow 0$ implying (a) $t(0)
\rightarrow 0$, (b) $t(0) \ll t(z > 1)$ as shown in figure \ref{figage}.
\subsection{Tracker}
\label{tracker}
\begin{figure*} \centering
\begin{center}
$\begin{array}{c@{\hspace{0.3in}}c}
\multicolumn{1}{l}{\mbox{}} &
        \multicolumn{1}{l}{\mbox{}} \\ [0.0cm]
\epsfxsize=1.95in
\epsffile{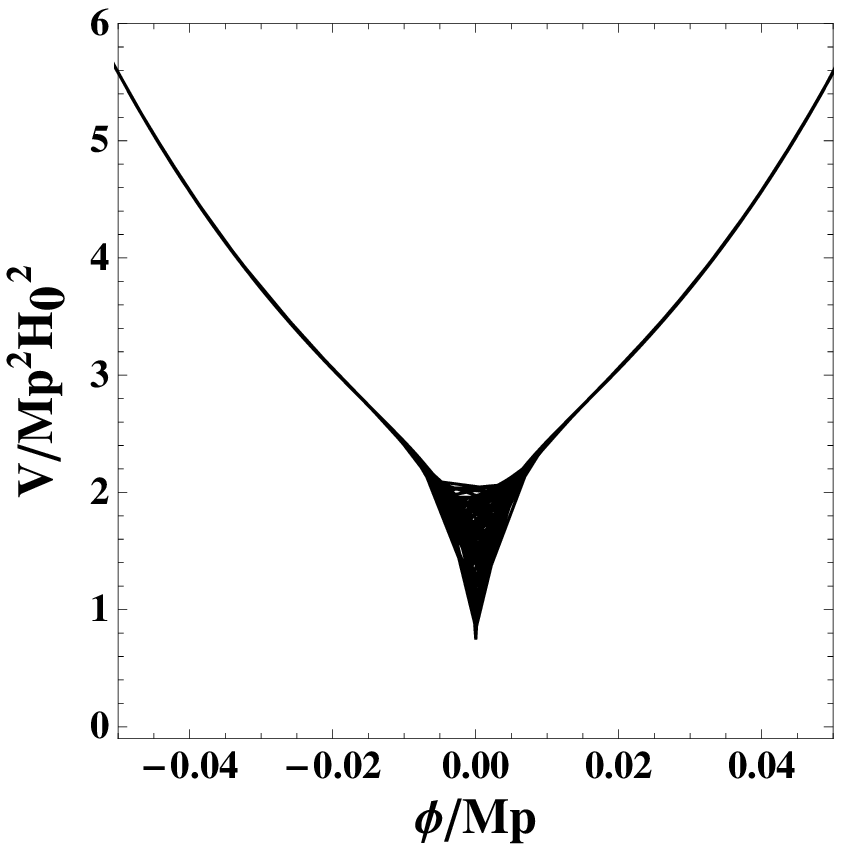} &
        \epsfxsize=2.07in
        \epsffile{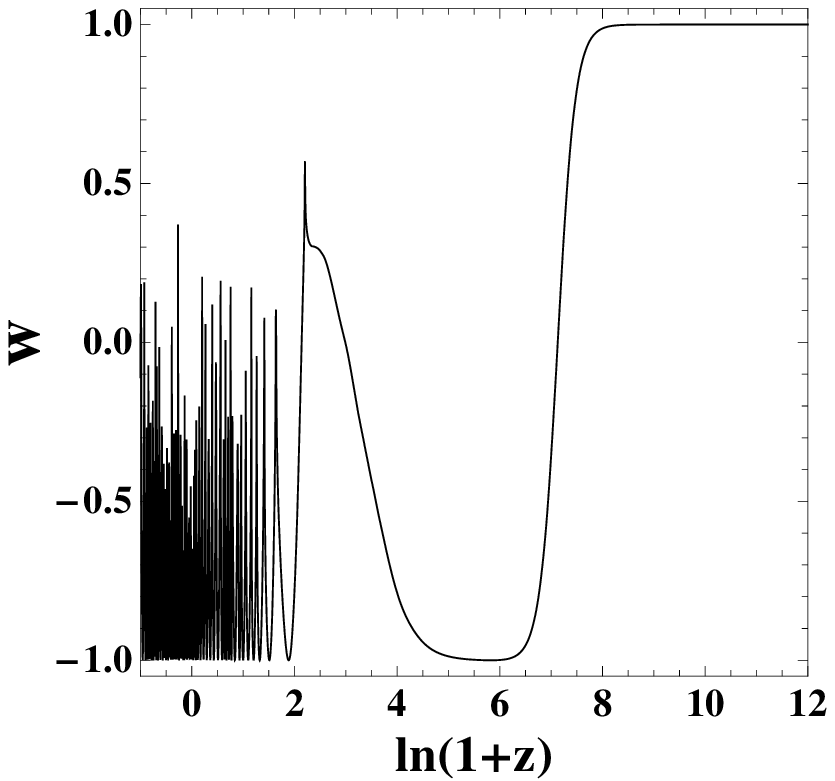}
\end{array}$
\end{center}
\caption{ This figure shows the evolution of potential (\ref{cosine}) and w versus field $\phi$ and redshift $z$ respectively. Figure shows that at late times the equation of state oscillates between zero and $-1$ such that the system spends most of the time around w$=-1$. We have taken $\Omega_{0m}=0.3$, $p=0.1$, $\tilde\alpha=200$ ($\alpha=p \tilde\alpha=20$) and $V_0=2.2~ M_{p}^2 H_0^2$.}
\label{figtracv}
\end{figure*}
\begin{figure*} \centering
\begin{center}
$\begin{array}{c@{\hspace{0.1in}}c c}
\multicolumn{1}{l}{\mbox{}} &
        \multicolumn{1}{l}{\mbox{}} \\ [0.0cm]
\epsfxsize=2.1in
\epsffile{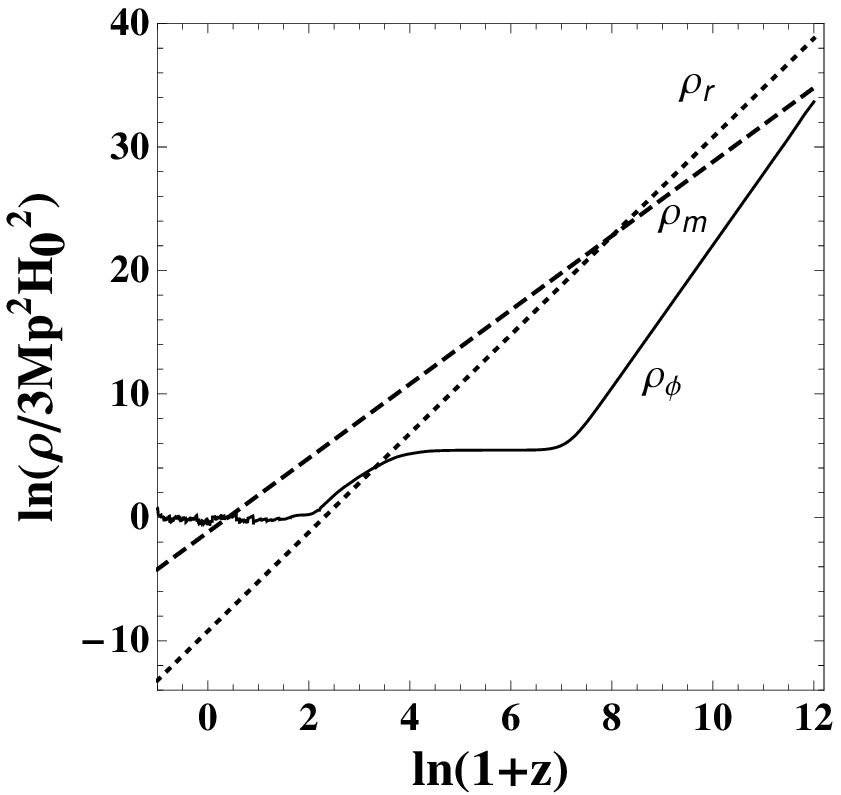} &
        \epsfxsize=2.0in
        \epsffile{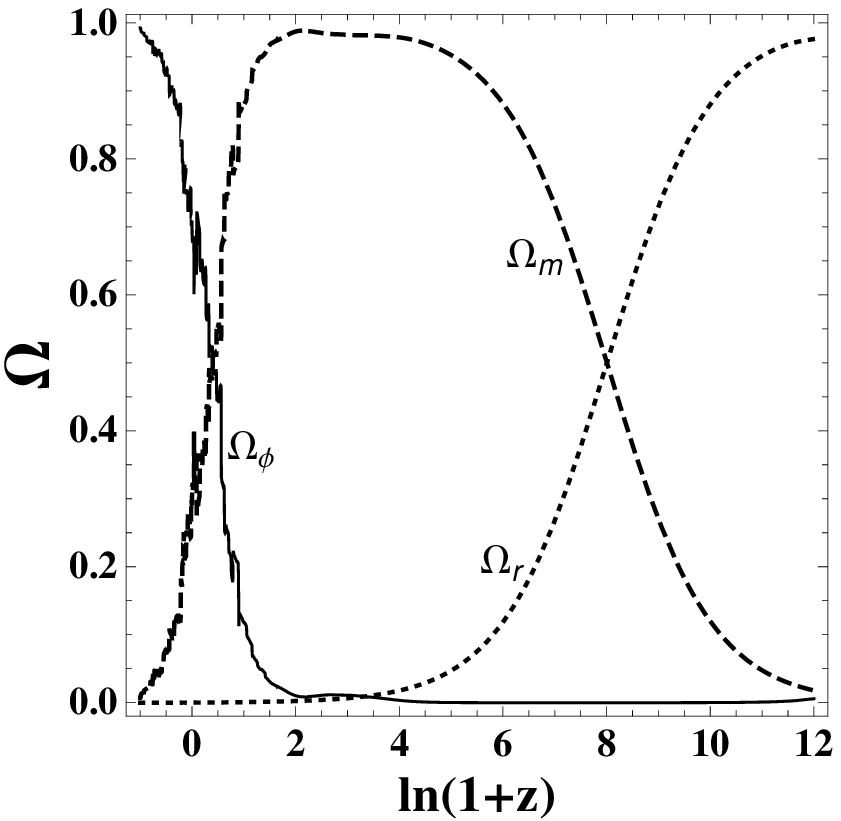} &
  \epsfxsize=2.01in
        \epsffile{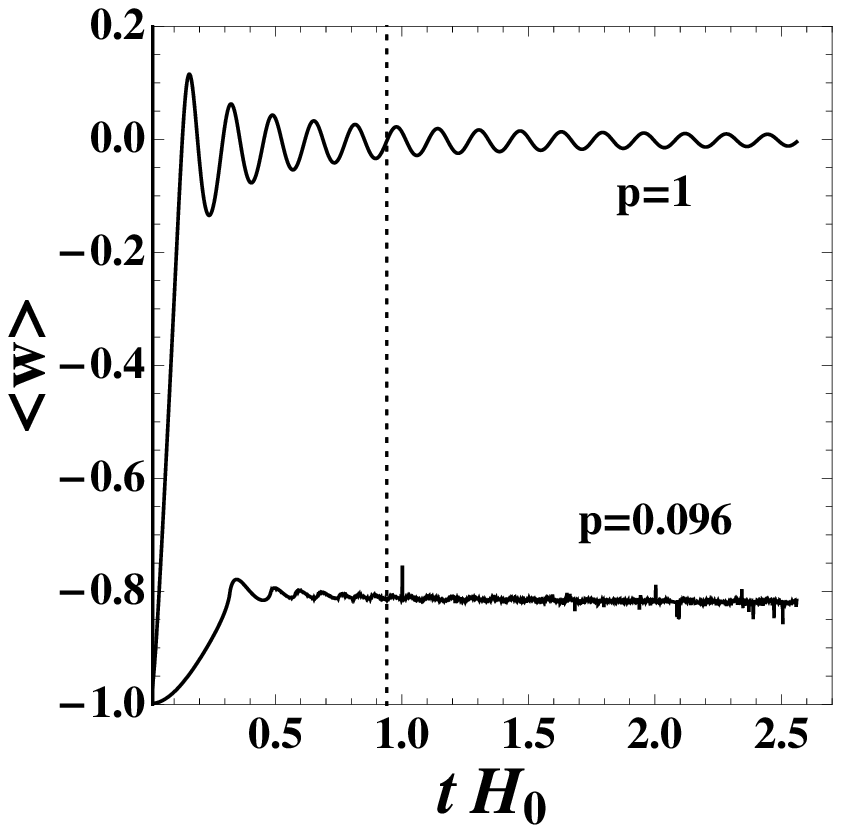}
\end{array}$
\end{center}
\caption{ This figure shows different plots for the potential (\ref{cosine}). The left and middle plots show the tracking behaviour of scalar field energy density and evolution of density parameter versus the redshift $z$. The dotted,
dashed and black lines correspond to 
 the radiation, matter and scalar field respectively. In both the plots the parameters are same as in figure \ref{figtracv}. The right plot shows the evolution of average equation of state $<$w$>$ versus cosmic time which agrees with $<$w$>$ = $\frac{p-1}{p+1}$, our plot corresponds to $p=1$, $\Omega_{0m}=0.319$, $\tilde\alpha=13$, $V_0=2.2~ M_{p}^2 H_0^2$ and $p=0.096$, $\Omega_{0m}=0.319$, $\tilde\alpha=37.59$, $V_0=2.2~ M_{p}^2 H_0^2$. The vertical dotted line designates the present epoch.}
\label{figtrac}
\end{figure*}
\begin{figure*} \centering
\begin{center}
$\begin{array}{c@{\hspace{0.1in}}c c}
\multicolumn{1}{l}{\mbox{}} &
        \multicolumn{1}{l}{\mbox{}} \\ [0.0cm]
\epsfxsize=1.95in
\epsffile{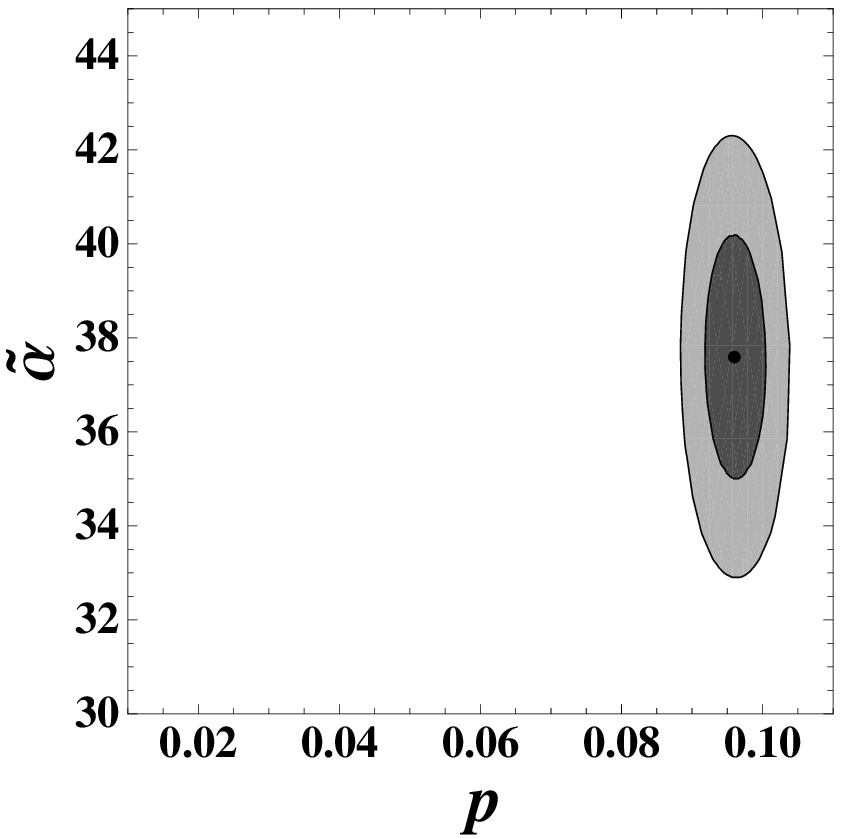} &
        \epsfxsize=2.05in
        \epsffile{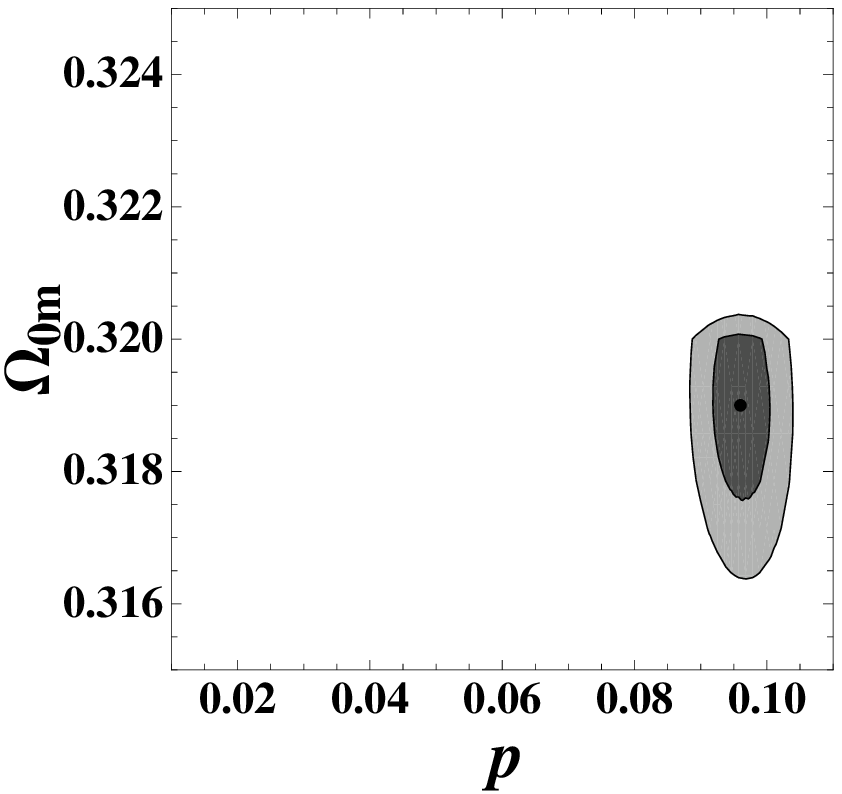} &
  \epsfxsize=2.05in
        \epsffile{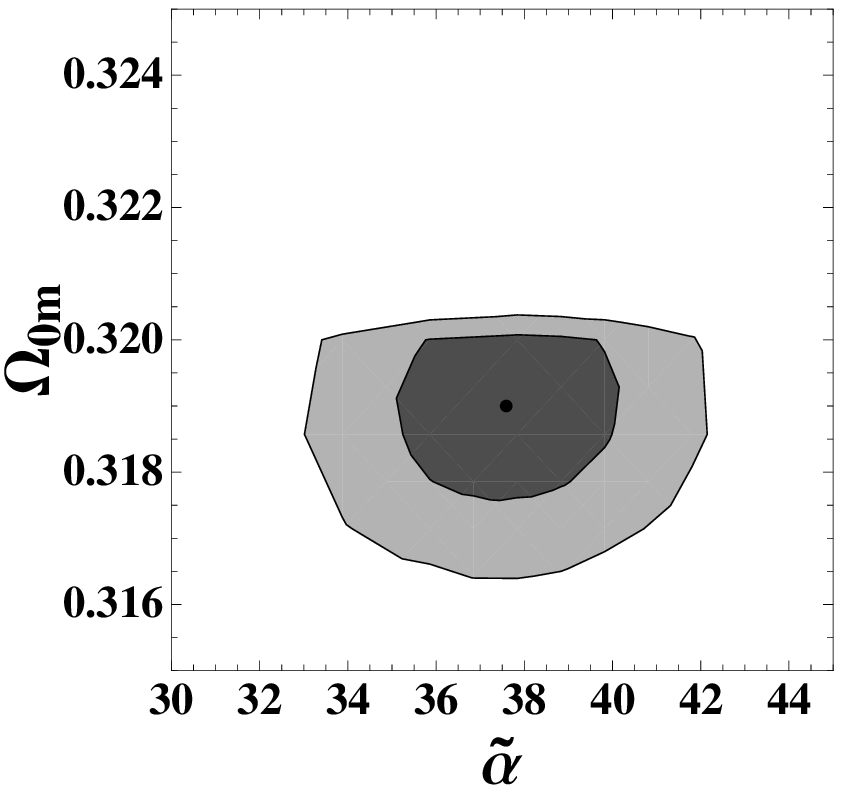}

\end{array}$
\end{center}
\caption{ This figure shows the 1$\sigma$ (dark shaded) and 2$\sigma$ (light shaded) likelihood contours for the potential (\ref{cosine}) with joint data (SN+Hubble+BAO+CMB); black dots designate the best fit value of the parameters which are found to be $\Omega_{0m} = 0.319$, $p = 0.096$ and $\tilde\alpha=37.59$.}
\label{figtraccont}
\end{figure*}
\begin{figure*} \centering
\begin{center}
$\begin{array}{c@{\hspace{0.3in}}c}
\multicolumn{1}{l}{\mbox{}} &
        \multicolumn{1}{l}{\mbox{}} \\ [0.0cm]
\epsfxsize=2.07in
\epsffile{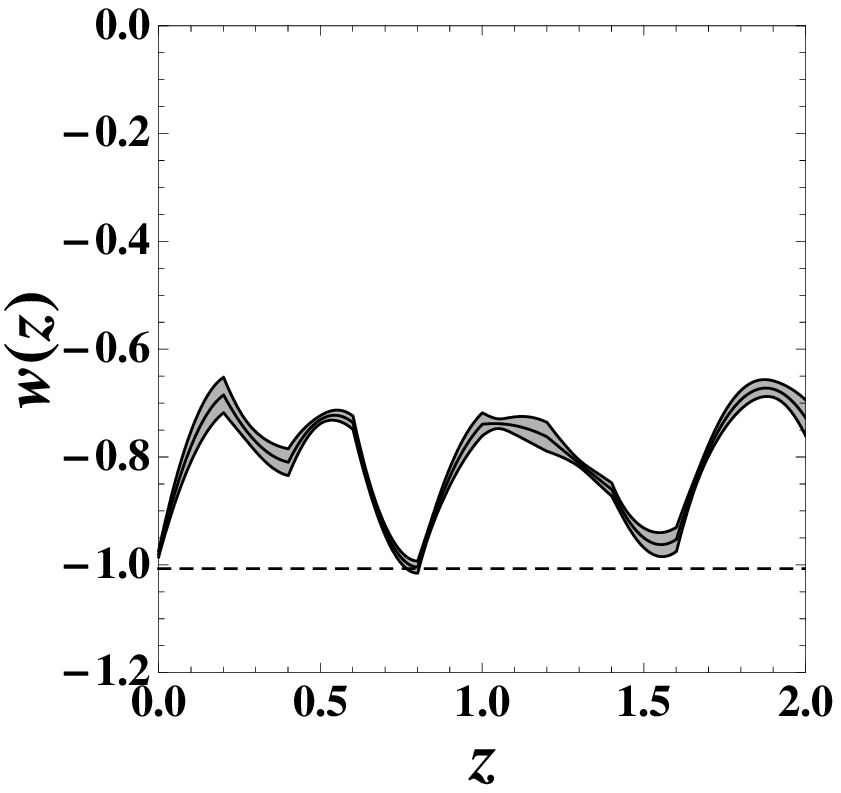} &
        \epsfxsize=2.02in
        \epsffile{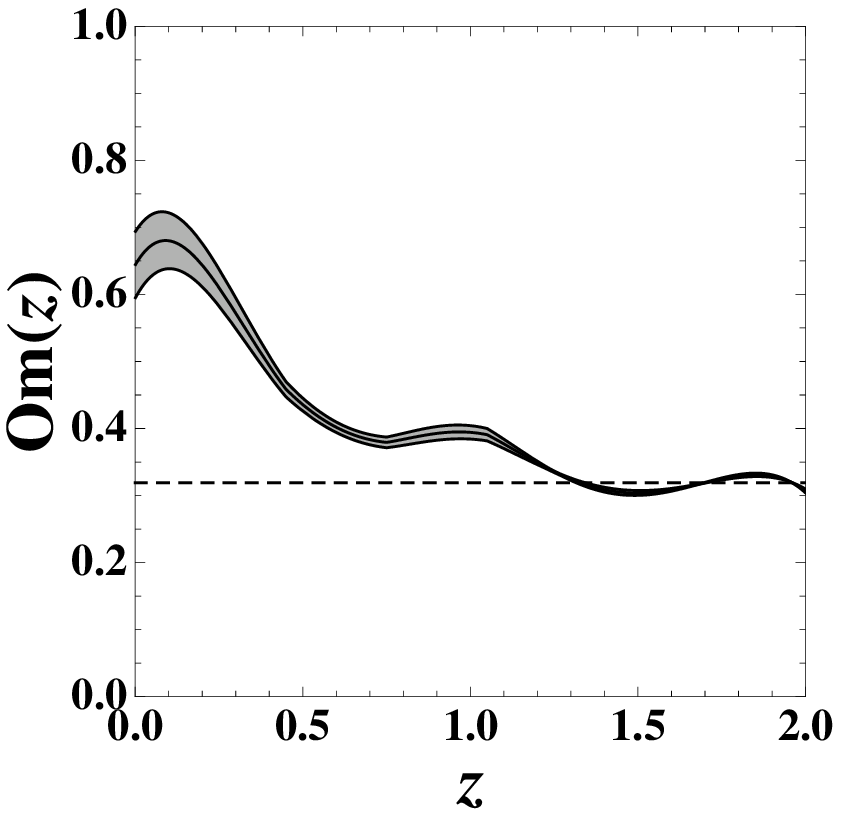}
\end{array}$
\end{center}
\caption{  This figure shows the evolution of $w(z)$ and $Om(z)$ versus the  redshift $z$ for the potential (\ref{cosine}). Equation of state exhibits oscillating behaviour which has also been manifested in $Om(z)$  plot. In both the plots, dashed line represents $\Lambda$CDM with $\Omega_{0m}=0.319$; solid (middle) lines inside shaded regions show best fitted behaviour and shaded regions show 1$\sigma$ confidence level. We have used joint data (SN+Hubble+BAO+CMB) in our analysis.}
\label{figtracom}
\end{figure*}
Scalar field models with tracker solutions are of
great importance in cosmology. A tracker solution is such that,
during radiation and matter era, the field mimics the background$-$
scaling regime and at late times it exits to late time cosmic acceleration. Once the model parameters
are fixed, the exit to dark energy behaviour remains independent for a wide variety of initial conditions. Here we shall
consider the following potential (Sahni \& Wang 2000): \be
V(\phi)=V_0\left[\cosh ({\tilde \alpha} \phi/M_p)-1\right]^p,~~~~0 <
~p < ~1/2 \label{cosine} \ee The above potential has asymptotic form
\be V(\phi)={V_0 \over 2^p} e^{p{\tilde\alpha}
\phi/M_p},~~~~~p\tilde\alpha  \vert\phi\vert /M_p\gg 1~,
\label{exppot} \ee and \be V(\phi)={V_0 \over
2^p}\left({\tilde\alpha} \phi \over M_p \right)^{2p},
~~~~~p\tilde\alpha  |\phi| /M_p \ll 1~, \label{ppot} \ee
where $p{\tilde \alpha} = \alpha$. The power law behaviour of
(\ref{cosine}) near the origin leads  to oscillations of $\phi$ when
it  approaches the origin. To have viable thermal history of the
Universe, we need to have $\Omega_{\phi}=
\frac{3(1+w_B)}{\alpha^2}$= constant$\leq 0.01$ (Ade et al. 2013)
during the radiation dominated era, which implies $\alpha \geq 20$
(here $w_B$ represents background equation of state). In this case,
the scalar field behaves like background matter, for most of the
history of Universe and, only at late times, it exits to dark
energy which demands for tracker solution.

The potential under consideration has peculiarity near the origin for generic values of $p$. We have noticed that
it is problematic to obtain oscillatory behaviour in the convex core of (\ref{cosine}), using the autonomous system of equations. In what follows, we shall use the dimensionless form of evolution equations which allows us to control the dynamics of field, beginning from scaling regime to oscillations about $\phi=0$. The equations of motion have the following form,
\begin{equation}
{\dot{a}^2(t) \over a^2(t)}={1 \over 3 M_{p}^2}  \left[\rho_{\phi}+\rho_m +\rho_r \right]
\label{friedman}
\end{equation}
\begin{equation}
\ddot{\phi}+ 3 {\dot{a} \over a}\dot{\phi}+V_{,\phi}(\phi)=0.
\label{evoleq}
\end{equation}
In order to investigate the dynamics described by equations (\ref{friedman}) and (\ref{evoleq}), it would be convenient to cast them as a system of first order equations
\begin{equation}
Y_1'=\frac{ Y_2}{h(Y_1,Y_2)} \,
\label{evol1d}
\end{equation}
\begin{equation}
Y_2'= -3Y_2-{1 \over h(Y_1, Y_2)}\Big[{d {\cal V}(Y_1) \over dY_1} \Big]
\label{evol2d}
\end{equation}
where
 \begin{equation}
 Y_1={\phi \over M_p},\quad Y_2={\dot{\phi} \over M_{p} H_{0}},\quad {\cal V}={ V(Y_1) \over M_p^2 H_0^2}
 \end{equation}
and prime denotes the derivative with respect to the variable $N=\ln(a)$. The function
$ h(Y_1, Y_2)$ is given as:
\begin{equation}
 h(Y_1, Y_2)=\sqrt{\left[{Y_2^2 \over 6}+ {{\cal V}(Y_1) \over 3} +{\Omega_{0m} e^{-3a}} +{\Omega_{0r} e^{-4a}}  \right]} \label{hubble}
\end{equation}
where $\Omega_{0m}$ and $\Omega_{0r}$ are the present energy density parameters of matter and radiation respectively.
We have numerically solved the evolution equations (\ref{evol1d}) $\&$  (\ref{evol2d});
Our  results are shown in figures \ref{figtracv}, \ref{figtrac}, \ref{figtraccont} and \ref{figtracom}.
Figure \ref{figtracv} shows the behaviour of the potential (\ref{cosine}) versus field $\phi$ and the equation of state parameter w versus the redshift $z$. As the field evolves from the steep region towards the
 origin, $\rho_\phi$ undershoots the background and field freezes for a while due to Hubble damping (see the 
 left and the middle plots of figure \ref{figtrac}). The field
 then approximately mimics the background before approaching the convex core of the potential, where oscillations set in. It is clear from the right plot of figure \ref{figtracv} that field oscillates most of the time near w$=-1$. The right plot of figure \ref{figtrac} shows average value of w versus the cosmic time for $p=1$ $\&$ $p=0.096$ which agrees with the analytical result, $<$w$>=p-1/p+1 $ at the attractor point.

In figure \ref{figtraccont}, all plots show the 1$\sigma$ (dark shaded) and 2$\sigma$ (light shaded) likelihood contours. Figure \ref{figtracom} shows the oscillating behaviour of equation of state and the corresponding behaviour of $Om$  inside 1$\sigma$ confidence level. The joint data
(SN+Hubble+BAO+CMB) was used for carrying out the observational analysis (see appendix A). We should note that the best fit value for the parameter $p$ is different from $p=0$ that would correspond to the case of cosmological constant. The latter
is reinforced by figure \ref{figtracom} which clearly shows, the model under consideration differs from $\Lambda$CDM
within  1$\sigma$ confidence level. In this case, we should further check for $\chi^2_{red}$. We find that the numerical value of $\chi^2_{red}=0.849713$ corresponding to the best fit value of $p$, is much smaller than one (see Table \ref{chi}). Thus, we can not conclude that the underlying model is preferred over $\Lambda$CDM.
In this case, we could not find slowing down effect, for any values of the model parameters.
\begin{table}
\caption{Comparison between $\chi_{red}^2$ values for the potential (\ref{cosine}) and $\Lambda$CDM. To obtain $\chi_{red}^2$ for the potential (\ref{cosine}), we vary $p$ and put the best fit values of the remaining parameters, $\Omega_{0m} = 0.319$ and $\tilde\alpha=37.59$. The bold values correspond to the best fit value of $p$ and $\Omega_{0m}$ for the potential (\ref{cosine}) and $\Lambda$CDM respectively.}
\label{chi}
\begin{center}
\begin{tabular}{c || cccccc}
\hline\hline
Potential (19)~~   & $\Lambda$CDM\\
$p$ ~~~~~ $\chi_{red}^2$ & ~~~$\Omega_{0m}$ ~~~~~ $\chi_{red}^2$\\
\\
\hline
0.01	~~~~~0.951237  & ~~~0.25	~~~~~0.956061\\
0.02	~~~~~0.973990  & ~~~0.26	~~~~~0.948286\\
0.03	~~~~~0.987693  & ~~~0.27	~~~~~0.943602\\
0.04	~~~~~0.992347  & ~~~{\bf 0.28}	~~~~~{\bf 0.94174}\\
0.05	~~~~~0.985540  & ~~~0.29	~~~~~0.942457\\
0.06	~~~~~0.972094  & ~~~0.30	~~~~~0.945545\\
0.07	~~~~~0.952011  & ~~~0.31	~~~~~0.950822\\
0.08	~~~~~0.888724  & ~~~0.32	~~~~~0.958119\\
0.09	~~~~~0.855366  & ~~~0.33    ~~~~~0.967293\\
{\bf 0.096}   ~~~~~{\bf 0.849713}  & ~~~0.34	~~~~~0.978213\\ 
0.10	~~~~~0.851936  & ~~~0.35	~~~~~0.990760      
\\
\hline\hline
\end{tabular}
\end{center}
\end{table}
\subsection{Phantom field with linear potential}
\label{phantom}
\begin{figure*} \centering
\begin{center}
$\begin{array}{c@{\hspace{0.3in}}c c}
\multicolumn{1}{l}{\mbox{}} &
        \multicolumn{1}{l}{\mbox{}} \\ [0.0cm]
\epsfxsize=2in
\epsffile{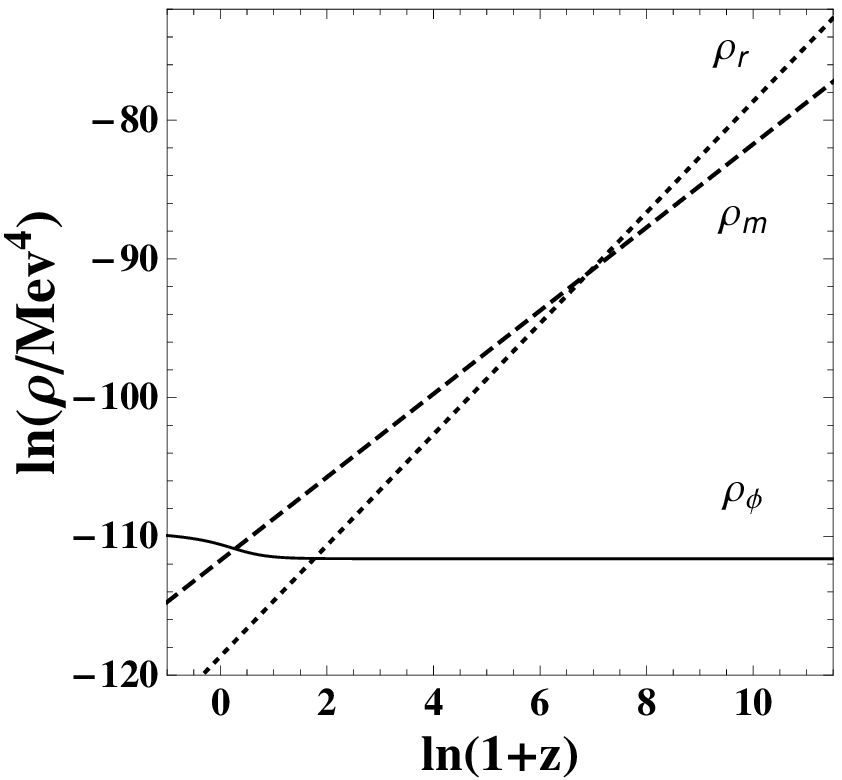} &
        \epsfxsize=2in
        \epsffile{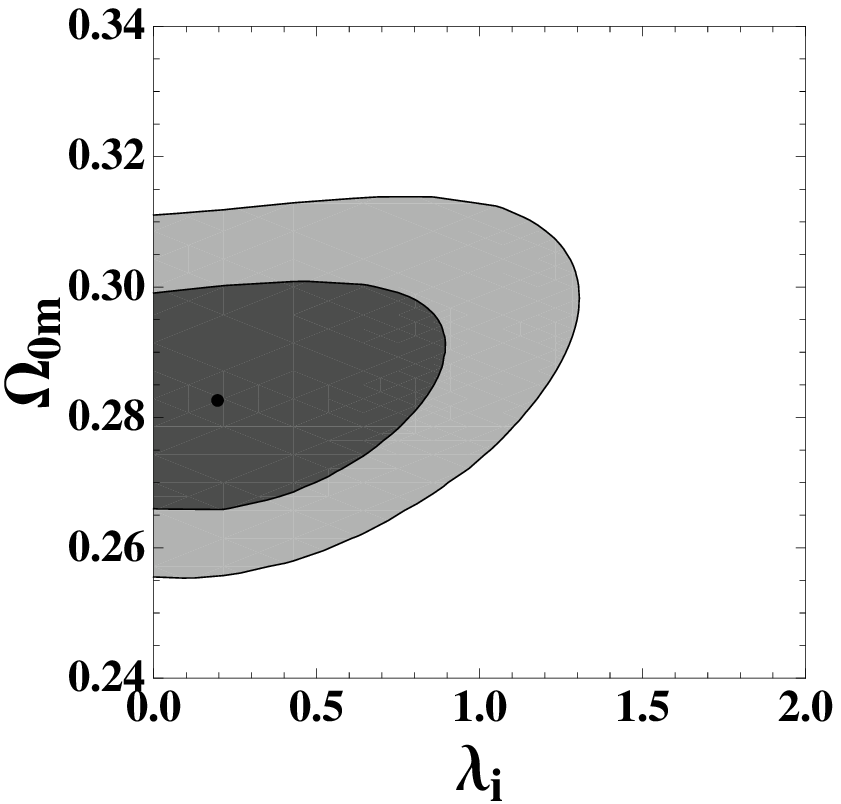}
\\
\multicolumn{1}{l}{\mbox{}} &
        \multicolumn{1}{l}{\mbox{}} \\ [0.0cm]
\epsfxsize=2in
\epsffile{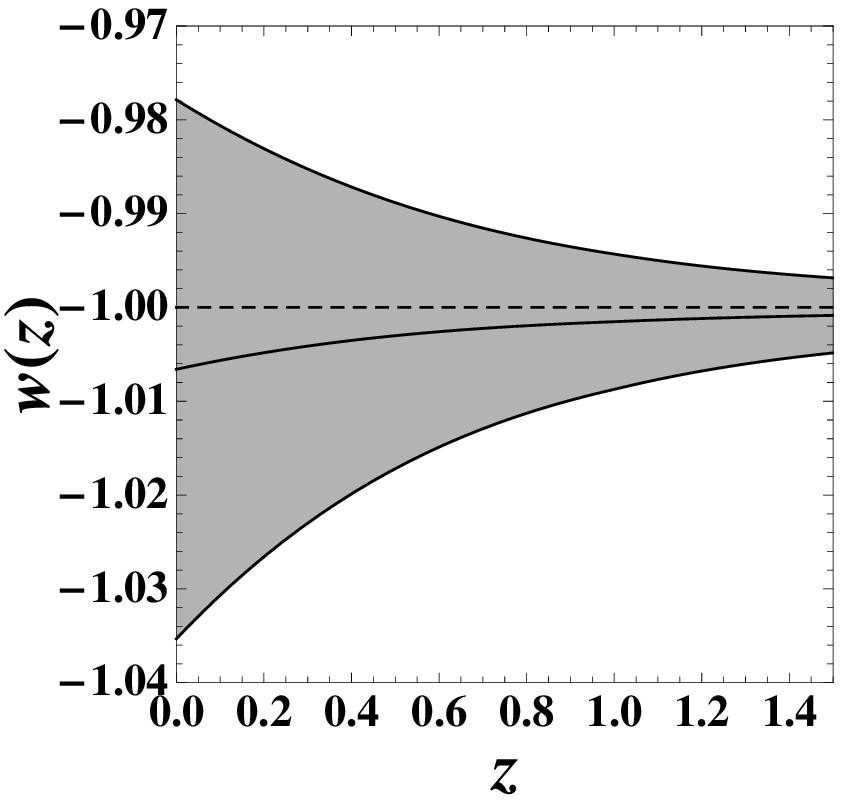} &
        \epsfxsize=1.96in
        \epsffile{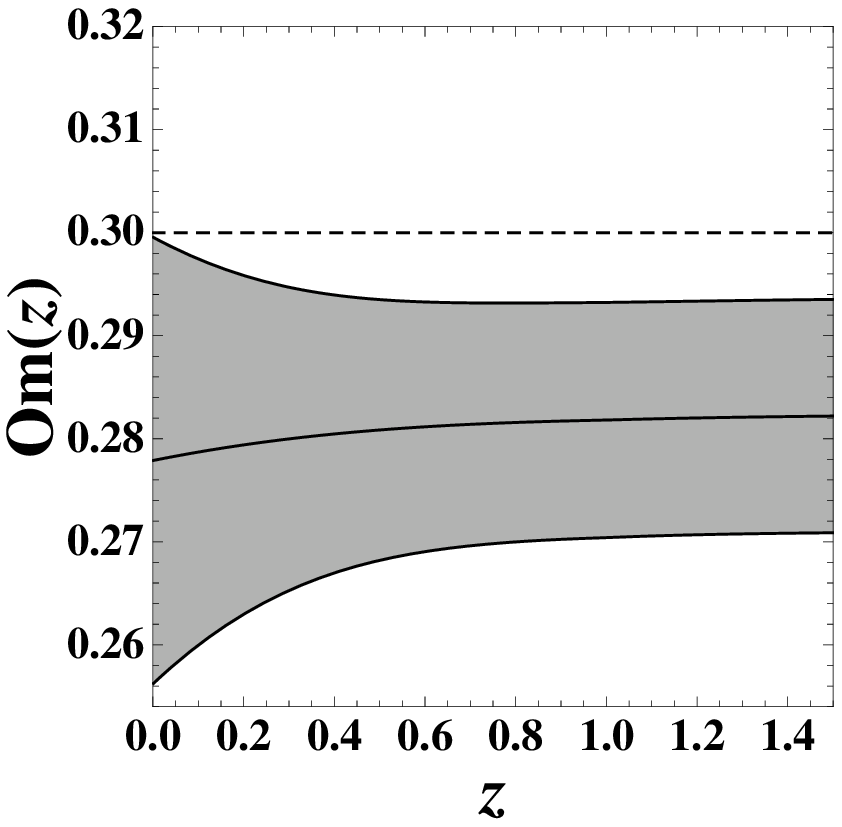}
\end{array}$
\end{center}
\caption{ This figure shows different plots for the linear potential. The
upper left plot shows the evolution of energy density versus the
redshift $z$. The dotted,
dashed and black lines correspond to 
 the
energy density of radiation, matter and phantom field respectively.
Initially, the energy density of phantom field is extremely
sub-dominant and remains to be so, for most of the period of
evolution.
 At late times, the field energy density catches up with the background, overtakes it, and
 starts growing ($w < -1$) and derives the current accelerated expansion of the Universe.
 The upper right plot shows the 1$\sigma$ (dark shaded) and 2$\sigma$ (light shaded) likelihood
 contours in the $\lambda_i - \Omega_{0m}$ plane
 (the black dot designates the best fit value of $\lambda_i$ and $\Omega_{0m}$, where $\lambda_i$ is the initial value of $\lambda$). The lower left and right plots show
  the evolution of $w(z)$ and $Om(z)$ versus the redshift $z$ respectively. $Om(z)$ has positive curvature.
  In both lower plots, the horizontal dashed line represents
  $\Lambda$CDM with $\Omega_{0m}=0.3$, solid (middle) line  inside
  shaded regions show best fitted behaviour and
  shaded regions show 1$\sigma$ confidence level. We have used joint data
   (SN+Hubble+BAO+CMB) in carrying out the analysis. }
\label{figphant}
\end{figure*}
Phantom dark energy with equation of state $w<-1$
can be achieved by introducing a negative kinetic energy term in the
action of the scalar field. By putting $\epsilon = -1$ in equations
(\ref{eq:phidd})
 and (\ref{eq:rhop}), we get equation of motion, energy density and pressure of phantom field
(Singh, Sami \& Dadhich 2003). In what follows, we shall examine the dynamic of a phantom field. In
this case, the Hubble parameter for spatially flat Universe can be
written as
\begin{eqnarray}
 H^2(z)&=&H_0^2 \Big{[} \Omega_{0r} (1+z)^4+\Omega_{0m} (1+z)^3  \nonumber\\
&& +  \Omega_{0\phi} \exp
\left( 3 \int_{0}^{z} \left[ 1 + w_{\phi}(z') \right]
\frac{dz'}{1+z'} \right) \Big{]},
\label{FDeq}
\end{eqnarray}
where $\Omega_{0r}$, $\Omega_{0m}$ and $\Omega_{0\phi}$ are the
present energy density parameters of radiation, matter and field
respectively, and $H_0$ designates the value of Hubble parameter at
the present epoch. We will be interested in the phantom dynamics
with a linear potential,
\be
V(\phi)=V_0 \phi,
\label{eq:phantpot} 
\ee
Numerically integrating the equations of motion, we find the field
energy density, equation of state and $Om$  for the said potential. The
results are shown in figure \ref{figphant}. The upper left plot shows
the evolution of energy density versus redshift $z$ whereas the
upper right plot shows the 1$\sigma$ (dark shaded) and 2$\sigma$ (light shaded)
likelihood contours in the $\lambda_i - \Omega_{0m}$ plane. The best fit values
 of the parameters are found to be $\lambda_i = 0.1959$ and $\Omega_{0m} = 0.2826$.
 The lower plots show the evolution of $w(z)$ and $Om(z)$. As seen in the figure,  $Om$
 has positive curvature for phantom field model which  distinguishes  phantom field model
 from zero-curvature ($\Lambda$CDM), for any current value of the matter density. The joint data i.e. SN+Hubble+BAO+CMB  was used for analysis (see appendix A).
\section{ non minimally coupled scalar field model}
\label{NMC}
\begin{figure*} \centering
\begin{center}
$\begin{array}{c@{\hspace{0.3in}}c}
\multicolumn{1}{l}{\mbox{}} &
        \multicolumn{1}{l}{\mbox{}} \\ [0.0cm]
\epsfxsize=2.04in
\epsffile{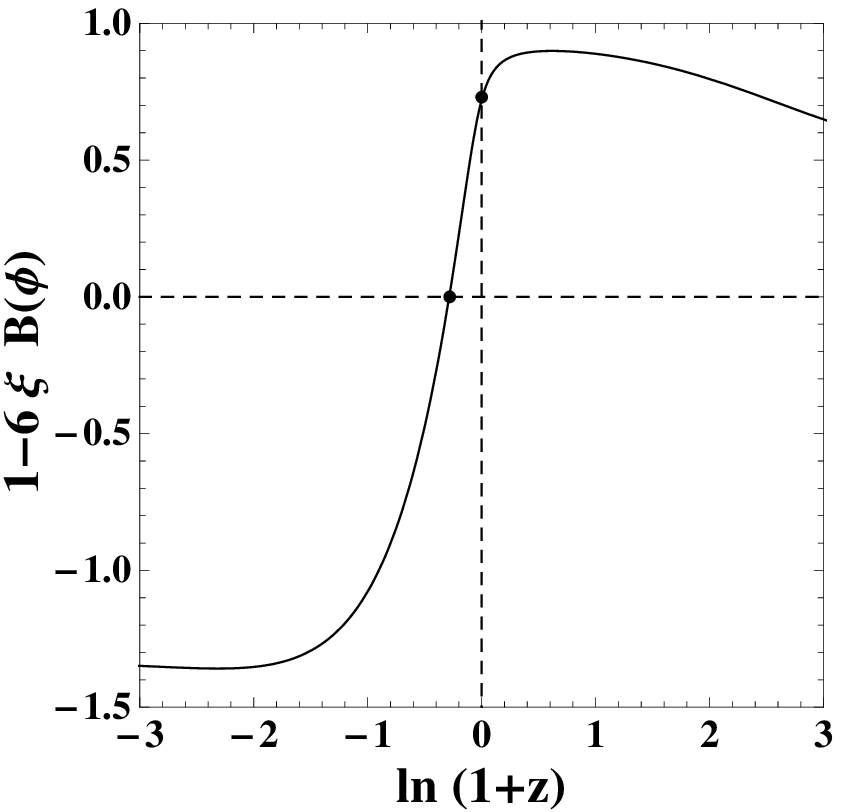} &
        \epsfxsize=2.03in
        \epsffile{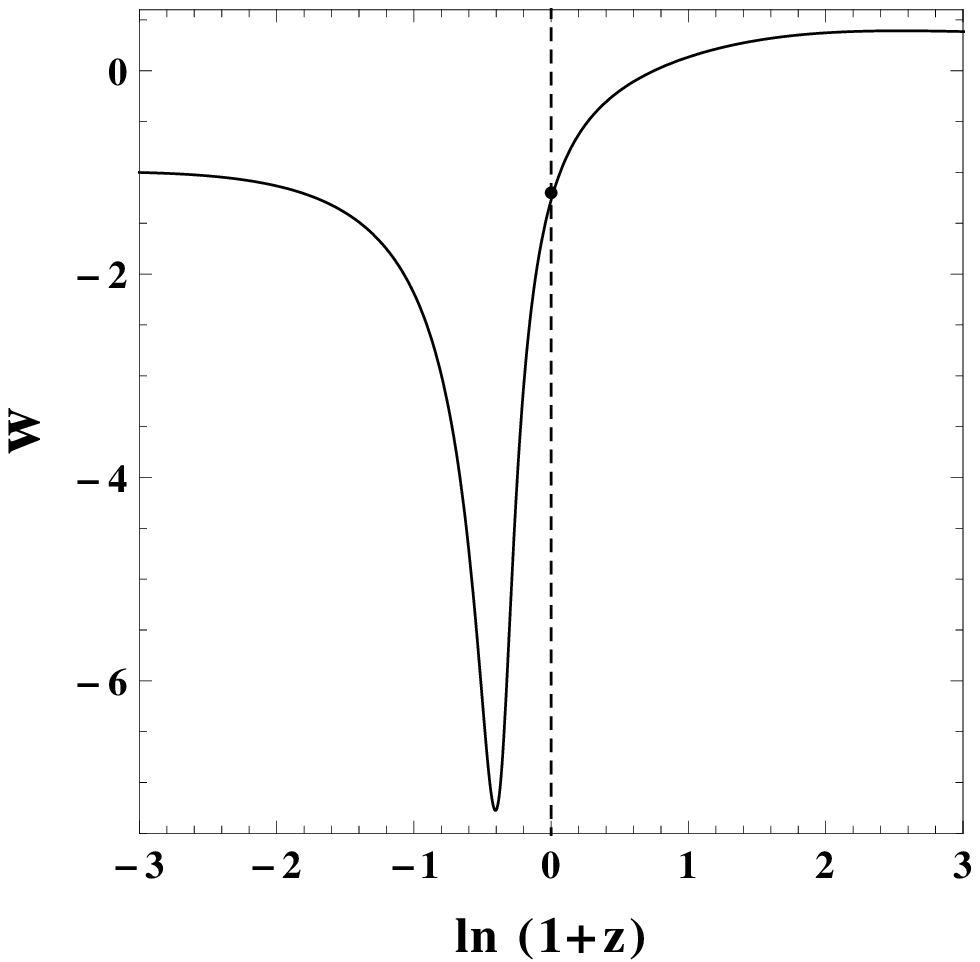}
\end{array}$
\end{center}
\caption{ This figure corresponds to the case of non minimal coupling
with $N=2$, $n=7$, $\xi= 0.2$ and $\Omega_{0m}=0.316$. The left and right plots show the evolution of $(1-6\xi B(\phi))$ and w versus the redshift $z$ respectively, where $G_{eff}=6/{8\pi(1-6\xi B(\phi))}$. The black dots ( point on the vertical dashed line) designate
 the present epoch which occurs in the regime of $G_{eff}>0$,
 the effective Newtonian constant changes sign thereafter in future.
  The point on the horizontal dashed line is the epoch where $G_{eff}$ turns negative.}
\label{wxinmc}
\end{figure*}
\begin{figure*} \centering
\begin{center}
$\begin{array}{c@{\hspace{0.1in}}c c}
\multicolumn{1}{l}{\mbox{}} &
        \multicolumn{1}{l}{\mbox{}} \\ [0.0cm]
\epsfxsize=2.06in
\epsffile{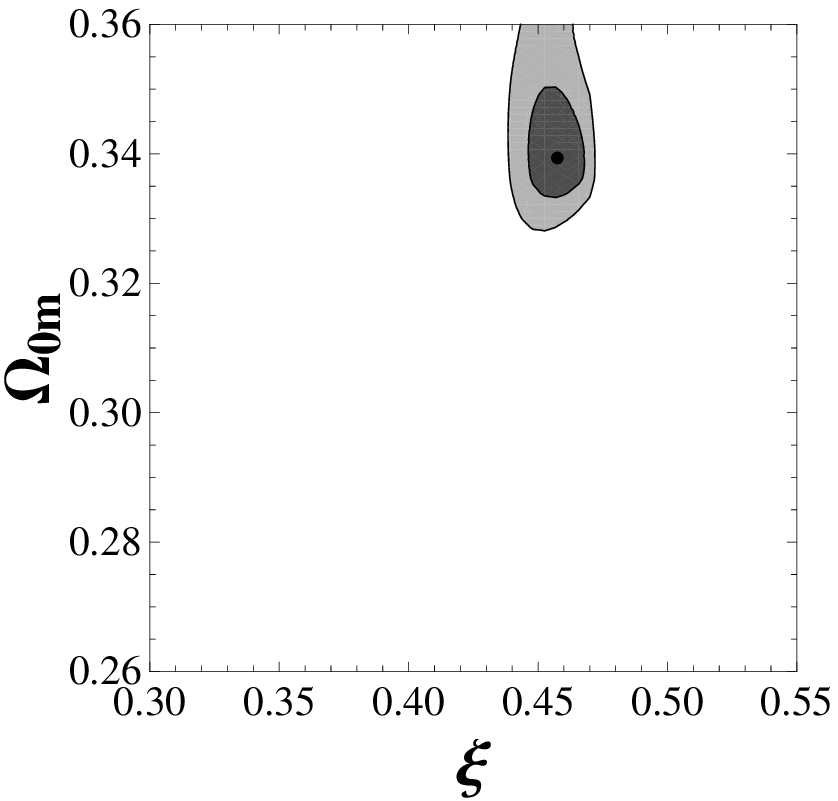} &
        \epsfxsize=2in
        \epsffile{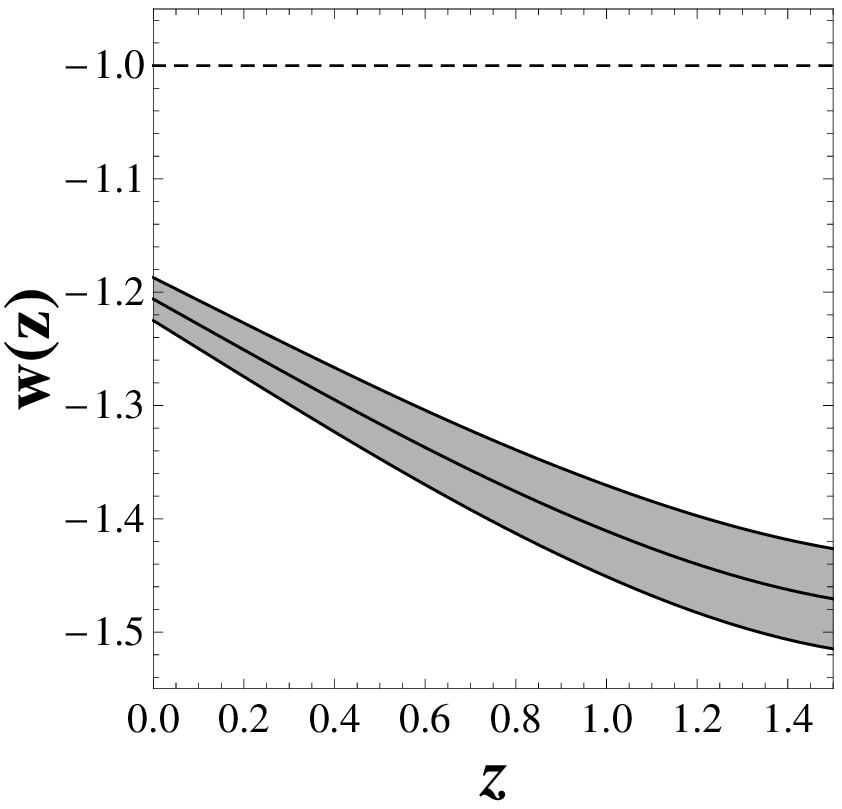} &
  \epsfxsize=2in
        \epsffile{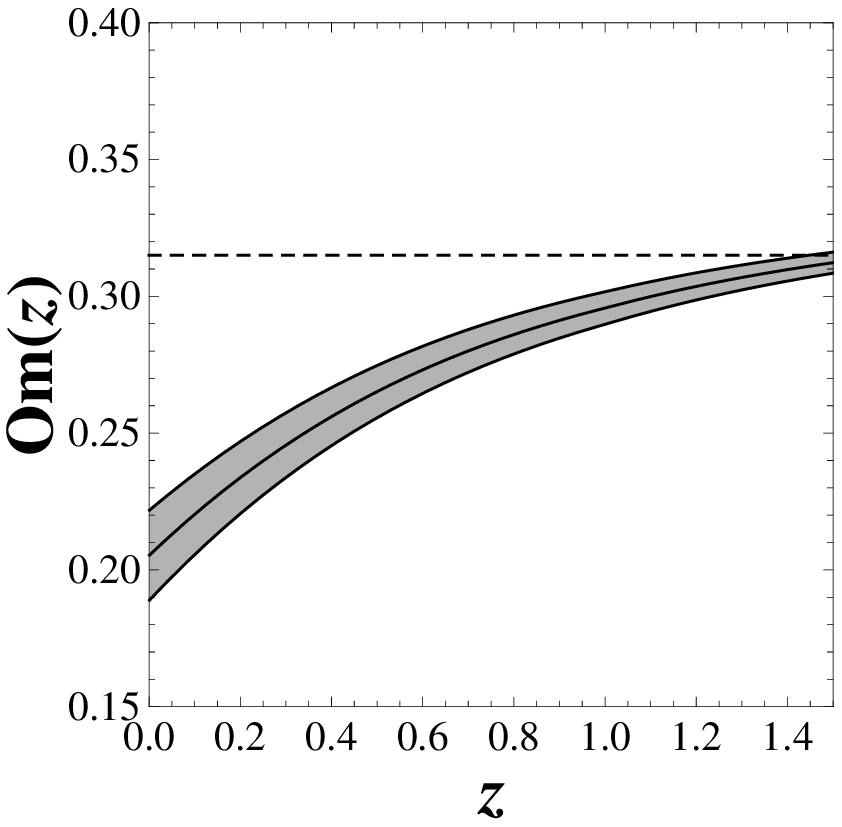}
\end{array}$
\end{center}
\caption{ This figure represents the non minimally coupled scalar field DE model with $N=2$, $n=7$; left plot shows the 1$\sigma$ (dark shaded) and 2$\sigma$ (light shaded) likelihood contours in the $\xi - \Omega_{0m}$ plane. The black dot corresponds to the best fit value. Evolution of w$(z)$ is shown in the middle plot. The right plot shows the evolution of $Om(z)$ versus the redshift $z$. In middle and right plots, solid (middle) line  inside shaded regions show best fitted behaviour and shaded regions show 1$\sigma$ confidence level; the horizontal dashed line represents $\Lambda$CDM with $\Omega_{0m}=0.315$. The SN+Hubble+BAO+CMB data has been used jointly in carrying out the analysis. }
\label{womnmc}
\end{figure*}
In this section, we revisit a non minimally
coupled scalar field model that allows to obtain a
transient phantom dark energy. Various features of
non minimally coupled scalar field system have been
investigated (Dent et al. 2013; Kamenshchik et al. 2013; Nozari \& Rashidi 2013; Aref'eva et al. 2014;  Kamenshchik et al. 2014;  Luo, Wu \& Yu 2014;  Pozdeeva \& Vernov 2014; Skugoreva, Toporensky \& Vernov 2014; Skugoreva, Saridakis \& Toporensky 2014). The action having non minimal coupling with scalar field is given as (Sami et al. 2012):
\be
\label{eq:Lagrangian}
S=\frac{1}{2}\int{\sqrt{-g}d^4x\Big{[} \frac{R} {\kappa}-(g^{\mu\nu}\phi_{\mu}\phi_{\nu}+ \xi R
B(\phi)+2V(\phi))\Big{]}}+S_M,
\ee
where $\kappa$=$8 \pi G$, $\xi$ is the dimensionless coupling constant and $S_M$ is the matter action.
The equations of  motion which are obtained by varying the action (\ref{eq:Lagrangian}) have the form (Sami et al. 2012):
\be
\label{eq:Friedphi}
H^2=\frac{\kappa}{3}\left(\frac{1}{2}{\dot
{\phi}}^{2}+V(\phi)+3\xi(H \dot {\phi} B'(\phi)+H^{2}B(\phi))+\rho
\right), \ee
\begin{eqnarray}
\label{eq:Friedphi2} R=\kappa \Big{(}-{\dot {\phi}}^{2} +4
V(\phi)+3\xi(3 H \dot{\phi}  B'(\phi)+\frac{R}{3}  B(\phi)\nonumber\\
  + {\dot {\phi}}^{2}B''(\phi)+\ddot {\phi} B'(\phi))+\rho(1-3\omega) \Big{)},
\end{eqnarray}
\begin{eqnarray}
\label{eq:KGphi}
&&\ddot {\phi}+3 H \dot {\phi}+\frac{1}{2}\xi R B'(\phi)+V'(\phi)=0,
\end{eqnarray}
where $R=6(2H^2+ \dot {H})$, is the Ricci Scalar and  $\rho$ is the energy density of the matter.

In this case (Sami et al. 2012), we have a varying effective Newtonian gravitational constant $G_{eff}$ which is a function of scalar field $\phi$. We found stationary points and their stability by considering $B(\phi) = {\phi}^N$ and $V(\phi) = V_0{\phi}^n$.
The de Sitter solution of interest to us, exists for $n<2N$  for which $G_{eff}={6}/{8\pi(1-6\xi
B(\phi))}=3(2N-n)/8\pi N$ is positive. In order to check the stability of the de Sitter solution, we
consider small perturbations, and get the system of equations (see appendix B)
\begin{equation}
\label{eq:matrix}
\left(\begin{array}{c}
\dot{\alpha}\\
\dot{\beta}\\
\dot{\gamma}
\end{array}\right)
=
\left(\begin{array}{ccc}
-\frac{A_1}{A_2}&-\frac{A_3}{A_2}&-\frac{A_4}{A_2}\\
 0&0&1\\
0&-\frac{A_5}{A_7}&-\frac{A_6}{A_7}
\end{array}\right)
\left(\begin{array}{c}
\alpha\\
\beta\\
\gamma
\end{array}\right)
\end{equation}
The numerical analysis of the system exhibits that the de Sitter solution is stable when $2N+1\leqslant n$ and for positive values of the coupling, $\xi >0$ (we assume $V_0=1$). Thus we display few cases in Table \ref{table:fixpoints} which show how the nature of the fixed points depends upon numerical values of $\xi$.
\begin{table}
\caption{ Nature of the fixed points (Sami et al. 2012):}
\begin{center}
\label{table:fixpoints}
\begin{tabular}{cccc}
\hline\hline
N & n & $\xi$ & nature of fixed points\\
\hline
2 & 5 & $0<\xi\leq 0.1333$ & saddle  \\
  &   & $0.1333<\xi\leq 0.2068$ & attractive focus  \\
  &   & $0.2069\leq \xi\leq 1$  & attractive node \\\\
2 & 7 & $0<\xi\leq 0.0952$      & saddle \\
  &   & $0.0952<\xi\leq0.3999$  & attractive focus\\
  &   & $0.4000\leq \xi\leq 1$  & attractive node \\\\
2 & 9 & $ 0<\xi \leq  0.0740$   & saddle  \\
  &   & $0.0740<\xi\leq 1$       & attractive focus \\\\
4 & 9 & $0<\xi<0.0009$          & saddle  \\
  &   & $0.0009<\xi\leq0.0014$  & attractive focus \\
  &   & $0.0015\le \xi\leq 1$   & attractive node \\\\
\hline\hline
\end{tabular}
\end{center}
\end{table}
As demonstrated by Sami et al. (2012), the de Sitter solution in this case occurs in the region of negative effective gravitational constant, thereby leading to a ghost dominated Universe in future and a transient quintessence (phantom) phase with $G_{eff}>0$ around the present epoch. Figure \ref{wxinmc} shows that before going to de Sitter point with $G_{eff}<0$, the equation of state passes through a phantom phase.
In order to obtain phantom ($w< - 1$) phase consistent with observation and $G_{eff}>0$ at present epoch, we adjust the model parameters as shown in figure \ref{wxinmc}.

Let us now apply  $Om$ diagnostic to the case under consideration
and examine the  curvature (slope) of $Om$. To this effect, we
investigate the evolution equations numerically in case of $N=2$
and $n=7$ for a viable range of parameters $\xi$ and $\Omega_{0m}$
($0.24\lesssim \Omega_{0m}\lesssim 0.36$ and $0.12\lesssim \xi
\lesssim 0.61$); the range of $\xi$ is dictated by stability
considerations. Our results are displayed in figure \ref{womnmc}, in
which the left plot shows the 1$\sigma$ (dark shaded) and 2$\sigma$
(light shaded) likelihood contours in the $\xi - \Omega_{0m}$ plane.
The best fit values of the parameters are, $\xi = 0.4574$ and
$\Omega_{0m} = 0.3393$. The middle plot shows the evolution of
w$(z)$ versus redshift $z$ whereas the right plot displays  the
evolution of $Om(z)$ versus redshift $z$. We find that $Om$ has
positive curvature (slope) for phantom equation of state which is a
generic feature of dark energy models with $w<-1$. The positive
curvature of phantom distinguishes non minimally coupled scalar
field model from zero curvature $\Lambda$CDM model, for any value of
the matter density as shown in figure \ref{womnmc}. We used
SN+Hubble+BAO+CMB joint data for carrying out the observational
analysis, see appendix A for details.
\section{Conclusions}
In this paper, we have investigated  scalar field models (including
the phantom case) using the $Om$ diagnostic. We have specifically
focused on models with power law potentials that lead to the slowing down of late time cosmic acceleration.
In case of quintessence with quadratic and quartic potentials, we have demonstrated that the slowing down
phenomenon takes place for $z\lesssim 0.4$. Figures \ref{figphi2} and \ref{figphi4} show that at late times, the field energy density starts decreasing $(w > -1)$ and exhibits matter/radiation like behaviour corresponding to the deceleration of expansion. Consequently, the equation of state of dark energy, $Om(z)$ and the deceleration parameter $q$  grow
  for redshift in the  interval $z \in (0, 0.4)$. This signifies that cosmic
    acceleration might have already peaked and that we are presently observing its slowing down.
    We find that quintessence models have negative curvature that differentiates
    these models from zero curvature
    ($\Lambda$CDM), for any given current value of the matter density as
    shown by the $Om$ plots in figures \ref{figphi2} and \ref{figphi4}. The best fit values of
    the model parameters for $\phi^2$ and $\phi^4$ potentials are found to
    be $\lambda_i=1.2$, $\Omega_{0m}=0.2657$ and
$\lambda_i=1.4415$, $\Omega_{0m}=0.2663$ respectively. In these
models, the best fit value of $\Omega_{0m}$ is always less than that for its
counter part i.e. $\Lambda$CDM, which compensates for the effect of
intermediate slowing down. We have also investigated a model with a
$cosh$ potential (\ref{cosine}), which has the tracking property. For
small values of $\phi$, the said potential mimics a power law
behaviour and gives rise to oscillations of $\phi$ near the origin.
The tracking behaviour of the scalar field energy density, the evolution
of density parameter, the oscillating behaviour of the equation of state
at late times and the corresponding behaviour of $Om$ ( within
1$\sigma$ confidence level ) are shown in figures \ref{figtrac} and
\ref{figtracom}. The field oscillations in the convex core around the origin are reflected
in the behaviour of w (see figure \ref{figtracv}), such that the average equation of state parameter
is given by $<$w$>=p-1/p+1$ (see figure \ref{figtrac}). The best fit
 values of the model parameters are, $p=0.096$, $\tilde\alpha=37.59$, $\Omega_{0m}=0.319$. In this case, the $\Lambda$CDM is clearly outside 2$\sigma$ confidence level as shown in figure \ref{figtraccont}. This is
 also clear from the figure \ref{figtracom} and it deserves a comment.
 In order to draw a final conclusion, we looked for $\chi^2_{red}$ and found that $\chi^2_{red}$ is much smaller than one. Thus we can not claim that the model under consideration is better than $\Lambda$CDM.
 
Next, we applied the  $Om$ diagnostic to phantom dark energy. In this case,
 we considered the phantom field with a linear potential. In figure
 \ref{figphant}, we have
 shown that  $Om$  has positive curvature that distinguishes
 the phantom dark energy from the zero curvature $\Lambda$CDM,
 for any current value of the matter density. The best fit values of the model parameters
 are, $\lambda_i=0.1959$, $\Omega_{0m}=0.2826$.
 We also examined a non minimally coupled scalar field model,
 which has a transient phantom behaviour. In this case again,
 $Om(z)$ has positive curvature  as shown in the figure \ref{womnmc}.
The best fit values of the model parameters are found to be $\xi=0.4574$, $\Omega_{0m}=0.3393$.
 It signifies that the positive curvature of $Om$ is a generic feature of phantom dark energy, be it transient or otherwise.

We conclude that given the present data, the $Om$ diagnostic can clearly
distinguish between scalar field models and $\Lambda$CDM and that in case of quadratic and quartic potentials, there
exists a specific region in the parameter space which could allow for the
slowing down of late time cosmic acceleration.
\section*{Acknowledgements}
We thank S. G. Ghosh and V. Soni for their constant encouragement throughout the
work. MS thanks  R. Gannouji, M. W. Hossain and Sumit Kumar for useful discussions. We also thank S. Ahmad and S. Rani for helping us in improving the manuscript.
\section*{Appendix A: Observational data analysis}
\label{AppendixA}
\begin{table*}
\caption{Values of $\frac{d_A(z_\star)}{D_V(Z_{BAO})}$ for distinct values of $z_{BAO}$.}
\begin{center}
\begin{tabular}{c||cccccc}
\hline\hline
~~~~~~~~~$z_{BAO}$~~  & ~~0.106~  & 0.2~& 0.35~ & 0.44~& 0.6~& 0.73~~\\
\hline
~~~~~~~~~ $\frac{d_A(z_\star)}{D_V(Z_{BAO})}$~~ &  ~~$30.95 \pm 1.46$~~~ & $17.55 \pm 0.60$~~~
& $10.11 \pm 0.37$ ~~~& $8.44 \pm 0.67$~~~ & $6.69 \pm 0.33$~~~ & $5.45 \pm 0.31$~~
\\
\hline\hline
\end{tabular}
\label{baodata}
\end{center}
\end{table*}
We put constraints on the model parameters using recent
observational data, namely Type Ia Supernovae, BAO, CMB and data
of Hubble parameter. The total $\chi^2$ for joint data is defined as
\begin{align}
\chi_{\rm tot}^2=\chi_{\rm SN}^2+\chi_{\rm BAO}^2+\chi_{\rm Hub}^2+\chi_{\rm CMB}^2 \,,
\label{chi_tot}
\end{align}
where the $\chi^2_i$ for each data set is evaluated as
follows: First, we consider the Type Ia supernova observation which is one of
the direct probes for the cosmological expansion. We use Union2.1
compilation data (Suzuki et al. 2012) of 580 data points. For this
case, one measures the apparent luminosity of the supernova
explosion from the photon flux received. In the present context, one
of the most  relevant cosmological quantity is luminosity distance
 $D_{L}(z)$ defined  as,
 \begin{equation}
D_L(z)=(1+z)
\int_0^z\frac{H_0dz'}{H(z')}.
\end{equation}
 Cosmologists often use the distance modulus $\mu$  defined as  $\mu=m - M=5 \log D_L+\mu_0$, where $m$ and $M$
are the apparent and absolute magnitudes of the Supernovae and
$\mu_0=5 \log\left(\frac{H_0^{-1}}{\rm Mpc}\right)+2 5$ is a
nuisance parameter which is marginalized. The corresponding $\chi^2$
is written as
\begin{align}
\chi_{\rm SN}^2(\mu_0,\theta)=\sum_{i=1}^{580}
\frac{\left[\mu_{th}(z_i,\mu_0,\theta)-\mu_{obs}(z_i)\right]^2}{
\sigma_\mu(z_i)^2}\,,
\end{align}
where $\mu_{obs}$, $\mu_{th}$  and $\sigma_{\mu}$ represents the observed, theoretical distance
modulus and uncertainty in the distance modulus respectively; $\theta$ represents any parameter of the particular
model. Eventually, marginalizing $\mu_0$ following Lazkoz, Nesseris \& Perivolaropoulos (2005), we get
\begin{align}
\chi_{\rm SN}^2(\theta)=A(\theta)-\frac{B(\theta)^2}{C(\theta)}\,,
\end{align}
where,
\begin{align}
&A(\theta) =\sum_{i=1}^{580}
\frac{\left[\mu_{th}(z_i,\mu_0=0,\theta)-\mu_{obs}(z_i)\right]^2}{
\sigma_\mu(z_i)^2}\,, \\
&B(\theta) =\sum_{i=1}^{580}
\frac{\mu_{th}(z_i,\mu_0=0,\theta)-\mu_{obs}(z_i)}{\sigma_\mu(z_i)^2}\,, \\
&C(\theta) =\sum_{i=1}^{580} \frac{1}{\sigma_\mu(z_i)^2}\,.
\end{align}
Next, we use BAO data of $\frac{d_A(z_\star)}{D_V(Z_{BAO})}$ (Eisenstein et al. 2005; Percival et al. 2010; Beutler et al. 2011; Blake et al. 2011;  Jarosik et al. 2011;  Giostri et al. 2012), where $z_\star \approx 1091$ is
the decoupling time,  $d_A(z)=\int_0^z \frac{dz'}{H(z')}$ is the co-moving angular-diameter
distance and  $D_V(z)=\left(d_A(z)^2\frac{z}{H(z)}\right)^{\frac{1}{3}}$ is the dilation scale.
Data required for this analysis is shown in Table~\ref{baodata}.

The $\chi_\mathrm{BAO}^2$ is defined as (Giostri et al. 2012),
\begin{equation}
 \chi_{\rm BAO}^2=X^T C^{-1} X\,,
\end{equation}
where,
\begin{equation}
X=\left( \begin{array}{c}
        \frac{d_A(z_\star)}{D_V(0.106)} - 30.95 \\
        \frac{d_A(z_\star)}{D_V(0.2)} - 17.55 \\
        \frac{d_A(z_\star)}{D_V(0.35)} - 10.11 \\
        \frac{d_A(z_\star)}{D_V(0.44)} - 8.44 \\
        \frac{d_A(z_\star)}{D_V(0.6)} - 6.69 \\
        \frac{d_A(z_\star)}{D_V(0.73)} - 5.45
        \end{array} \right)\,,
\end{equation}
and $C^{-1}$ is the inverse covariance matrix given by Giostri et al. (2012).

We then use the observational data on Hubble parameter as recently
compiled by Farooq \& Ratra (2013) in the redshift range $0.07\leq z \leq 2.3 $. The sample contains
 28 observational data points of $H(z)$. These values are given in Table \ref{hubble}. To complete the
  data set, we take the latest and most precise measurement of the Hubble constant $H_0$
  from PLANCK 2013 results (Ade et al. 2013). To apply the data of Hubble parameter on our models, we
  work with the normalized Hubble parameter, $h=H/H_0$.

The $\chi^2$ for the normalized Hubble parameter is defined as,
\begin{align}
\chi_{\rm Hub}^2(\theta)=\sum_{i=1}^{29}
\frac{\left[h_{\rm th}(z_i,\theta)-h_{\rm obs}(z_i)\right]^2}{
\sigma_h(z_i)^2}\,,
\end{align}
where, $h_{\rm obs}$ and $h_{\rm th}$ are the  observed and theoretical values of
the normalized Hubble parameter respectively.\\
Also,
\be
\sigma_h = \left( \frac{\sigma_H}{H}+\frac{\sigma_{H_0}}{H_0} \right) h,
\label{errorh}
\ee
where $\sigma_H$  and $\sigma_{H_0}$ is the error in $H$ and ${H_0}$ respectively.

Finally, we apply CMB shift parameter $R=H_0 \sqrt{\Omega_{m0}}
\int_0^{1089}\frac{dz'}{H(z')}$. The corresponding $\chi_{\rm
CMB}^2$ can be written as,
\begin{align}
 \chi_{\rm CMB}^2(\theta)=\frac{(R(\theta)-R_0)^2}{\sigma^2}\,,
\end{align}
where, $R_0=1.725 \pm 0.018$ ( Komatsu et al. 2011).
\begin{table}
\caption{$H(z)$ measurements (in unit [$\mathrm{km\,s^{-1}Mpc^{-1}}$]) and their errors (Farooq \& Ratra 2013).}
\begin{center}
\label{hubble}
\begin{tabular}{cccc}
\hline\hline
~$z$ & ~~$H(z)$ &~~ $\sigma_{H}$ & ~~ Reference\\
0.070&~~    69&~~~~~~~  19.6&~~ Zhang et al. 2012\\
0.100&~~    69&~~~~~~~  12&~~   Simon et al. 2005\\
0.120&~~    68.6&~~~~~~~    26.2&~~ Zhang et al. 2012\\
0.170&~~    83&~~~~~~~  8&~~    Simon et al. 2005\\
0.179&~~    75&~~~~~~~  4&~~    Moresco et al. 2012\\
0.199&~~    75&~~~~~~~  5&~~    Moresco et al. 2012\\
0.200&~~    72.9&~~~~~~~    29.6&~~ Zhang et al. 2012\\
0.270&~~    77&~~~~~~~  14&~~   Simon et al. 2005\\
0.280&~~    88.8&~~~~~~~    36.6&~~ Zhang et al. 2012\\
0.350&~~    76.3&~~~~~~~    5.6&~~  Chuang \& Wang 2012\\
0.352&~~    83&~~~~~~~  14&~~   Moresco et al. 2012\\
0.400&~~    95&~~~~~~~  17&~~   Simon et al. 2005\\
0.440&~~    82.6&~~~~~~~    7.8&~~  Blake et al. 2012\\
0.480&~~    97&~~~~~~~  62&~~   Stern et al. 2010\\
0.593&~~    104&~~~~~~~ 13&~~   Moresco et al. 2012\\
0.600&~~    87.9&~~~~~~~    6.1&~~   Blake et al. 2012\\
0.680&~~    92&~~~~~~~  8&~~    Moresco et al. 2012\\
0.730&~~    97.3&~~~~~~~    7.0&~~   Blake et al. 2012\\
0.781&~~    105&~~~~~~~ 12&~~   Moresco et al. 2012\\
0.875&~~    125&~~~~~~~ 17&~~   Moresco et al. 2012\\
0.880&~~    90&~~~~~~~  40&~~  Stern et al. 2010\\
0.900&~~    117&~~~~~~~ 23&~~   Simon et al. 2005\\
1.037&~~    154&~~~~~~~ 20&~~   Moresco et al. 2012\\
1.300&~~    168&~~~~~~~ 17&~~   Simon et al. 2005\\
1.430&~~    177&~~~~~~~ 18&~~   Simon et al. 2005\\
1.530&~~    140&~~~~~~~ 14&~~  Simon et al. 2005\\
1.750&~~    202&~~~~~~~ 40&~~  Simon et al. 2005\\
2.300&~~    224&~~~~~~~ 8&~~    Busca et al. 2013\\

\hline\hline
\end{tabular}
\end{center}
\end{table}
\section*{Appendix B}
For de Sitter solution, substituting $\dot{H}=\dot{\phi}=\ddot{\phi}=\rho=0$ and $R=12{H_0}^2$ with $B(\phi) = {\phi}^N$ and $V(\phi) = V_0{\phi}^n$  in equations (\ref{eq:Friedphi}), (\ref{eq:Friedphi2}) and (\ref{eq:KGphi}), we obtain
\begin{eqnarray}
\label{eq:b14} {H_0}^2(1-6\xi {\phi_0}^N)&=&2 V_0 {\phi_0}^n,\nonumber\\
 6 {H_0}^2\xi N {\phi_0}^{N-1}+V_0 n {\phi_0}^{n-1}&=&0 ,
\end{eqnarray}
which gives
\begin{eqnarray}
\label{eq:b15} H_0^2=-\frac{V_0 n{\phi_0}^{n-N}}{6\xi  N},~~~
{\phi_0}^N=\frac{n}{6\xi (n-2N)}.
\end{eqnarray}
The de Sitter solution exists provided that $n<2N$, and $G_{eff}={6}/{8\pi(1-6\xi
B(\phi))}=3(2N-n)/8\pi N$ is positive only if $n<2N$.
     The system of equations  (\ref{eq:Friedphi}), (\ref{eq:Friedphi2}) and (\ref{eq:KGphi}) for $\rho=0$
      then  reduces to,
\begin{eqnarray}
\label{eq:ds1} \dot H (1-6\xi
B+9{\xi}^2{B'}^2)&=&3{\dot{\phi}}^2(\xi B''-1)\nonumber\\
&& -3\xi B'(4H \dot{\phi}+V'+6\xi B' H^2),\nonumber\\
\ddot{\phi}(1-6\xi B+9\xi^2{B'}^2)&=&-(3H \dot{\phi}+V')(1-6\xi
B)\nonumber\\
&& -\xi B'(-3\dot{\phi}^2+12V+9\xi(3H \dot{\phi} B'\nonumber\\
&& +{\dot{\phi}}^2
B'')).
\end{eqnarray}
To check the stability of the de Sitter, we
consider small perturbations $\mu$ $\&$ $\nu$ around this
background: $H=H_0+\mu$ and $\phi={\phi}_0+\nu$ in the dynamical
system (\ref{eq:ds1}) which gives us the evolution equations
for perturbations
\begin{eqnarray}
\label{eq:ds2}
&&\mu(36{\xi}^2  N^2 H_0 {{\phi}_0}^{2N-2})\nonumber\\
&& +\dot{\mu}(1-6\xi  {{\phi}_0}^N+9{\xi}^2 N^2{{\phi}_0}^{2N-2}) \nonumber\\
&& +\nu(3\xi  V_0 Nn(N+n-2){{\phi}_0}^{N+n-3}\nonumber\\
&& +36{\xi}^2  N^2(N-1){H_0}^2{{\phi}_0}^{2N-3})\nonumber\\
&& +\dot{\nu}(12\xi  N H_0 {{\phi}_0}^{N-1})
=0,
\end{eqnarray}
\begin{eqnarray}
\label{eq:ds3} &&\nu(V_0
n(n-1){{\phi}_0}^{n-2}(1-6\xi{\phi_0}^N)
\nonumber\\
&& -6\xi  V_0
Nn{{\phi}_0}^{N+n-2}+12\xi  V_0
N(N+n-1){{\phi}_0}^{N+n-2})\nonumber\\
&&+\dot{\nu}(3H_0 (1-6\xi {{\phi}_0}^N)+27{\xi}^2 N^2
H_0{{\phi}_0}^{2N-2}) \nonumber\\
&& +\ddot{\nu}(1-6\xi {{\phi}_0}^N+9{\xi}^2
N^2{{\phi}_0}^{2N-2})=0.
\end{eqnarray}
Equations (\ref{eq:ds2}) and (\ref{eq:ds3}) can be put in a
simple form by introducing the following notations
\begin{eqnarray}
A_1&=&36{\xi}^2  N^2 H_0 {{\phi}_0}^{2N-2},\nonumber\\
A_2&=&1-6\xi {{\phi}_0}^N+9{\xi}^2 N^2{{\phi}_0}^{2N-2},\nonumber\\
A_3&=&3\xi  V_0 Nn(N+n-2){{\phi}_0}^{N+n-3} \nonumber\\
&& +36{\xi}^2  N^2(N-1){H_0}^2{{\phi}_0}^{2N-3},\nonumber\\
A_4&=&12\xi  N H_0 {{\phi}_0}^{N-1},\nonumber\\
A_5&=&V_0 n(n-1){{\phi}_0}^{n-2}(1-6\xi {{\phi}_0}^N)-6\xi  V_0 Nn{{\phi}_0}^{N+n-2}\nonumber\\
&& +12\xi  V_0 N(N+n-1){{\phi}_0}^{N+n-2},\nonumber\\
A_6&=&3H_0 (1-6\xi {{\phi}_0}^N)+27{\xi}^2 N^2 H_0{{\phi}_0}^{2N-2},\nonumber\\
A_7&=&1-6\xi {{\phi}_0}^N+9{\xi}^2 N^2{{\phi}_0}^{2N-2}
\end{eqnarray}
Next, by using, $\alpha=\mu$, $\beta=\nu$, $\gamma=\dot{\nu}$,
the system of equations (\ref{eq:ds2}) and (\ref{eq:ds3})
acquires a simple form,
\begin{eqnarray}
&&A_1 \alpha +A_2 \dot{\mu} +A_3 \beta +A_4 \gamma =0,\\
&&A_5 \beta +A_6 \gamma +A_7 \ddot{\nu}=0.
\end{eqnarray}
Taking derivative of $\alpha$, $\beta$ and $\gamma$ with respect to time we get the system of equations,
\begin{equation}
\label{eq:matrix}
\left(\begin{array}{c}
\dot{\alpha}\\
\dot{\beta}\\
\dot{\gamma}
\end{array}\right)
=
\left(\begin{array}{ccc}
-\frac{A_1}{A_2}&-\frac{A_3}{A_2}&-\frac{A_4}{A_2}\\
 0&0&1\\
0&-\frac{A_5}{A_7}&-\frac{A_6}{A_7}
\end{array}\right)
\left(\begin{array}{c}
\alpha\\
\beta\\
\gamma
\end{array}\right)
\end{equation}

\end{document}